\documentclass[a4paper,11pt]{article}
\pdfoutput=1

\usepackage{epsfig,psfrag,graphics,verbatim}
\usepackage{graphicx}% Include figure files
\usepackage{epstopdf}
\usepackage{dcolumn}% Align table columns on decimal point
\usepackage{bm}% bold math
\usepackage{slashed}
\usepackage{color}
\usepackage{amssymb}
\usepackage{amsmath}
\usepackage{float}
\usepackage{placeins}

\usepackage{jcappub} % for details on the use of the package, please
\usepackage[T1]{fontenc} % if needed

\title{\boldmath Internal bremsstrahlung signatures in light of direct dark matter searches}

\author[a]{Mathias Garny,}
\author[b]{Alejandro Ibarra,}
\author[b]{Miguel Pato}
\author[b]{and Stefan Vogl}

\affiliation[a]{Deutsches Elektronen-Synchrotron DESY, Notkestra\ss{}e 85, 22603 Hamburg, Germany}
\affiliation[b]{Physik-Department T30d, Technische Universit\"at M\"unchen, James-Franck-Stra\ss{}e, 85748 Garching, Germany}

\emailAdd{mathias.garny@desy.de}
\emailAdd{ibarra@tum.de}
\emailAdd{miguel.pato@tum.de}
\emailAdd{stefan.vogl@tum.de}

\abstract{Although proposed long ago, the search for internal bremsstrahlung signatures has only recently been made possible by the excellent energy resolution of ground-based and satellite-borne gamma-ray instruments. Here, we investigate thoroughly the current status of internal bremsstrahlung searches in light of the results of direct dark matter searches and in the framework  of a minimal  mass-degenerate scenario consisting of a Majorana dark matter    particle that couples to a fermion and a scalar  via a Yukawa coupling. The upper limits on the annihilation cross section set by Fermi-LAT and H.E.S.S.~extend uninterrupted from tens of GeV up to tens of TeV and are rather insensitive to the mass degeneracy in the particle physics model. In contrast, direct searches are best in the moderate to low mass splitting regime, where XENON100 limits overshadow Fermi-LAT and H.E.S.S.~up to TeV masses if dark matter couples to one of the light quarks. In our minimal scenario we examine carefully the prospects for GAMMA-400, CTA and XENON1T, all planned to come online in the near future, and find that: (a) CTA and XENON1T are fully complementary, with CTA most sensitive to multi-TeV masses and mass splittings around 10\%, and XENON1T probing best small mass splittings up to TeV masses; and (b) current constraints from XENON100 already preclude the observation of any spectral feature with GAMMA-400 in spite of its impressive energy resolution, unless dark matter does not couple predominantly to light quarks. Finally, we point out that, unlike for direct searches, the possibility of detecting thermal relics in upcoming internal bremsstrahlung searches requires, depending on the concrete scenario, boost factors larger than 5-10.}

\begin{document}

\vspace{-0.5cm}
\begin{flushright}
\hfill{{\small DESY 13-115}  }\\
\hfill{{\small TUM-HEP 896/13 }}
\end{flushright}

\maketitle
\flushbottom

\section{Introduction}

\par If dark matter (DM) is constituted by weakly interacting massive particles (WIMPs) thermally produced in the early universe, the self-annihilations of dark matter particles in the Milky Way will produce fluxes of antimatter, neutrinos or gamma-rays which could be observed at the Earth as an excess over the expected astrophysical backgrounds. In many cases, however, the observation of such an excess cannot be automatically attributed to dark matter annihilations, due to the limited understanding of the mechanisms involved in producing the background fluxes. For example, it has been argued that the emission of electron-positron pairs in pulsar magnetospheres could generate an excess in the positron fraction over the expected values from secondary production \cite{Harding,Atoian:1995ux,Chi:1995id,Hooper:2008kg}, and similarly for the hadronic interactions of cosmic rays undergoing acceleration in supernova remnants, which could produce an excess in the antiproton-to-proton fraction \cite{Blasi:2009bd}. A remarkable exception would be the observation of a sharp gamma-ray spectral feature, which can be produced in dark matter annihilations \cite{Srednicki:1985sf,Rudaz:1986db,Bergstrom:1988fp,Bergstrom:1989jr,Flores:1989ru} but not by any (known) astrophysical mechanism (cf.~\cite{Aharonian:2012cs} for a possible exception), and thereby can be searched for very efficiently in the sky without requiring a precise determination of the background (see e.g.~\cite{Aharonian:2003xh,Bergstrom:2004cy,Bergstrom:2005ss,Bringmann:2007nk,Bringmann:2008kj,Abdo:2010nc,Vertongen:2011mu,Bringmann:2012vr,Weniger:2012tx, Ibarra:2012dw,Ackermann:2012qk,Bringmann:2012ez,Abramowski:2013ax, Fermi-LAT:2013uma}).

\par In this paper we will concentrate on scenarios where the dark matter particle couples to a light fermion via a Yukawa coupling. As shown in \cite{Bringmann:2007nk}, the three-body dark matter annihilation into a fermion-antifermion pair and a photon produces, in a process dubbed internal bremsstrahlung (IB), a gamma-ray spectrum with a pronounced feature close to the kinematic endpoint which resembles a distorted gamma-ray line. Searches for the spectral feature from internal bremsstrahlung have been conducted in Fermi-LAT \cite{Bringmann:2012vr} and H.E.S.S.~\cite{Abramowski:2013ax} data allowing to set fairly stringent limits on the annihilation cross section, which in some scenarios lie just one or two orders of magnitude above the expected value for a thermal relic. Future gamma-ray telescopes with enhanced energy resolution (such as GAMMA-400 \cite{Galper:2012fp} or DAMPE \cite{dampe}) or with larger effective areas (such as H.E.S.S.~II \cite{hesssite}, MAGIC-II \cite{Tridon2010437}, or CTA \cite{Consortium:2010bc}) will continue closing in over the next years on the search for the spectral feature of internal bremsstrahlung.

\par This dark matter scenario also predicts the generation of an antiproton flux, the scattering of dark matter particles with nuclei and the existence of events with jet(s) and large amounts of missing energy in high energy proton-proton collisions. Therefore, measurements of the cosmic antiproton-to-proton fraction, direct dark matter searches and collider experiments provide complementary limits on the parameter space of dark matter scenarios leading to internal bremsstrahlung, which might even preclude the observation of a gamma-ray feature in future instruments. Note that, since internal bremsstrahlung is most pronounced when dark matter interacts with the Standard Model via a light mediator, the corresponding phenomenology is distinct from the effective theory approach \cite{Goodman:2010qn,Rajaraman:2011wf,Fox:2011pm,Chu:2012qy,Frandsen:2012db,Cotta:2012nj,Rajaraman:2012fu, Gustafsson:2013gca} in various respects that we discuss in detail below. See \cite{Bergstrom:2010gh,Asano:2011ik,Lee:2012ph,Kopp:2013mi,Jackson:2013pjq,Liu:2013gba} for some recent works on the interplay of direct and collider searches with gamma-ray line searches featuring light mediators.

\par In this paper we aim at studying the complementarity of limits that direct and indirect dark matter searches impose on the parameter space of a toy model producing internal bremsstrahlung (as well as gamma-ray lines), in order to assess the prospects to observe gamma-ray spectral features at the future gamma-ray telescopes GAMMA-400 and CTA. We also briefly discuss the relevance of recent collider searches. The paper is organised as follows. In section \ref{ppmodel} we briefly review some properties of the class of mass-degenerate dark matter models under consideration. In section \ref{searches} we discuss the various experimental searches and provide some details on our analyses. We discuss current limits and their complementarity as well as future prospects in section \ref{results}, and then present our conclusions in section \ref{conclusions}.

\section{Particle physics model and internal bremsstrahlung}\label{ppmodel}

\par We consider a minimal extension of the Standard Model (SM) that generates an internal bremsstrahlung signal and where the dark matter is constituted  by a Majorana fermion $\chi$ which interacts with a SM fermion and a scalar $\eta$ via a Yukawa interaction with coupling constant $f$. We assume $\chi$ to be a singlet under the SM gauge group while $\eta$ can be either a singlet or a triplet under $SU(3)_c$. Then, the Lagrangian is given by:
\begin{align}
  {\cal L}={\cal L}_{\rm SM}+{\cal L}_{\chi}+{\cal L}_\eta+ {\cal L}_{\rm int}
  \;.
\end{align} 
Here, ${\cal L}_{\rm SM}$ is the SM Lagrangian, whereas ${\cal L}_{\chi}$ and ${\cal L}_\eta$ are the parts of the Lagrangian involving just the new fields $\chi$ and $\eta$. They are given, respectively, by
\begin{align}
  \begin{split}
    {\cal L}_\chi&=\frac12 \bar \chi^c i\slashed {\partial} \chi
    -\frac{1}{2}m_\chi \bar \chi^c\chi\;, \; \text{and}\\ {\cal L}_\eta&=(D_\mu
    \eta)^\dagger  (D^\mu \eta)-m_\eta^2 \eta^\dagger\eta \;,
  \end{split}
\end{align}
where $D_\mu$ denotes the usual covariant derivative. 

\par Lastly, ${\cal L}_{\rm int}$ describes the interaction of the dark matter particle with the SM. In order to avoid flavour changing neutral currents we assign a flavour quantum number to $\eta$. Concretely, we will restrict our discussion to dark matter particles interacting just with the right-handed $u$-quarks, $b$-quarks or  muons (the analysis for couplings to left-handed fermions is completely analogous). Thus the interaction term in the Lagrangian is given by
\begin{align}
  {\cal L}_{\rm int} &= - f \bar \chi \psi_R \eta+{\rm h.c.} \;,
\label{eq:singlet-eR}
\end{align}
with $\psi=u,\, b \, \text{or}\, \mu $. Furthermore, the model allows the term $H^\dagger H\eta^\dagger\eta$, where $H$ denotes the SM Higgs doublet. For couplings of order one or less this term is not relevant for the direct nor the indirect  dark matter detection, hence we will neglect it in the rest of the paper.

\par The interactions in this model allow tree-level annihilations into pairs of SM fermions $\psi \bar \psi$. The cross section can be expanded in partial waves $\sigma v = a + b v^2 + O(v^4)$ ,  the s-wave contribution being \cite{EllisPhys.Lett.B444:367-3721998}
\begin{align}
  (\sigma v)_\text{2-body}^\text{$s$-wave}= \frac{f^4 N_c}{32\pi m_\chi^2}
  \frac{m_{\psi}^2}{m_\chi^2}\frac{1}{(1+\mu)^2}\;,
  \label{eqn:sv2s}
\end{align}
where $N_c$ is a colour factor and $\mu\equiv(m_\eta/m_\chi)^2$ is a parameter that measures the mass splitting between the dark matter particle $\chi$ and the mediator $\eta$. The s-wave contribution is helicity-suppressed, as apparent from Eq.~\eqref{eqn:sv2s} by the presence of the factor $m_{\psi}^2/m_{\chi}^2$, while the p-wave contribution is also suppressed due to the small relative velocity of the dark matter particles in the galactic halo, $v\sim 10^{-3}$. Therefore, the $2\rightarrow 2$ annihilation cross section is very small and higher order processes might become relevant.

\par Concretely, there are two final states, a priori suppressed, which can become relevant or even dominant in this class of scenarios. The first final state arises from the annihilation into a fermion-antifermion pair with the associated emission of a vector boson, $\psi \bar \psi V$, where $V$ can either be a photon $\gamma$, a gluon $g$ or a weak gauge boson \cite{Bergstrom:1989jr,Bringmann:2007nk,Flores:1989ru,Drees:1993bh,Bell:2011eu,Bell:2011if,Bell:2012dk,Ciafaloni:2011sa,Ciafaloni:2011gv,Ciafaloni:2012gs,DeSimone:2013gj,Garny:2011cj,Garny:2011ii,Asano:2011ik,Fukushima:2012sp}. Although the $2\rightarrow 3$ process is suppressed by a phase space factor and by an additional coupling constant, it can have a larger cross section than the $2\rightarrow 2$ process due to the large suppression of the latter by the factor $m_{\psi}^2/m_{\chi}^2$ or by $v^2$. The total s-wave annihilation cross section into two massless fermions and one photon is given by \cite{Bell:2011if,Bringmann:2007nk}
\begin{align}
  (\sigma v)_\text{3-body}
  \simeq  &\frac{\alpha_\text{em} f^4 N_c Q_\psi^2}{64\pi^2m_\chi^2}
   \left\{ 
  (\mu+1) \left[ \frac{\pi^2}{6}-\ln^2\left( \frac{\mu+1}{2\mu} \right)
  -2\text{Li}_2\left( \frac{\mu+1}{2\mu} \right)\right]  \right. \nonumber \\
   &+ \left. \frac{4\mu+3}{\mu+1}+\frac{4\mu^2-3\mu-1}{2\mu}\ln\left(
  \frac{\mu-1}{\mu+1} \right)
  \right\}\;.
  \label{eqn:sv3}
\end{align}
For dark matter particles coupling to light fermions and small values of $\mu$, the cross section of this process can be more than one order of magnitude larger than the two body annihilation into a fermion-antifermion pair. The full expression for the differential cross section can be found e.g.~in \cite{Garny:2011ii}. The second final state of interest arises from the annihilation into two vector bosons, $V V$, which is generated at the quantum level and which is loop- but not helicity-suppressed. Again, in some instances the helicity suppression can be stronger than the loop suppression, making the annihilation process $\chi \chi\rightarrow VV $ relevant. The cross section for this process  is rather lengthy, therefore we refer the reader to \cite{Bergstrom1997}. For $V=\gamma$, the two final states produce a line or a line-like feature in the gamma-ray energy spectrum which could be detected in present or future gamma-ray telescopes.

\begin{figure*}[htp]
\centering
\includegraphics[width=0.49\textwidth]{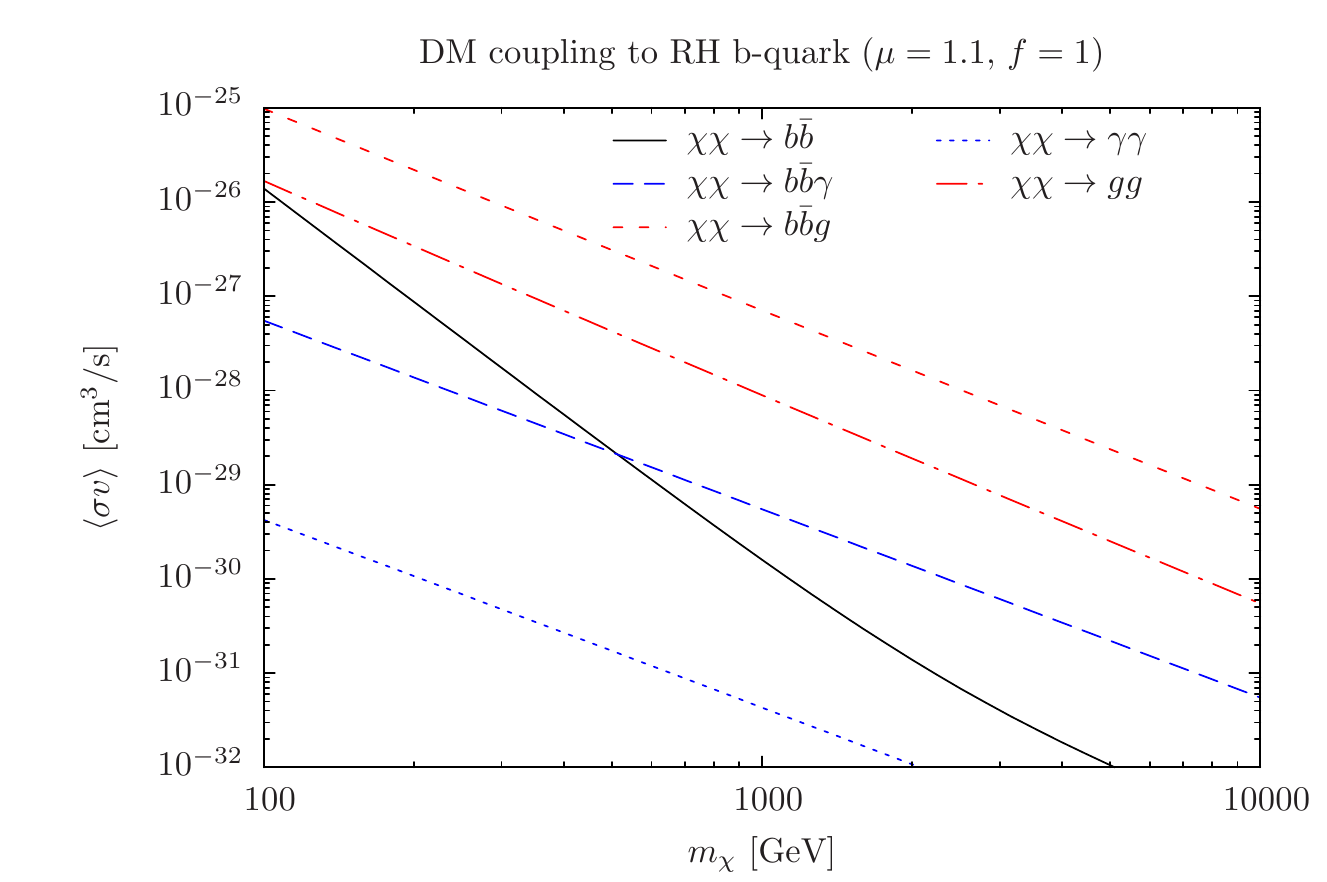}
\includegraphics[width=0.49\textwidth]{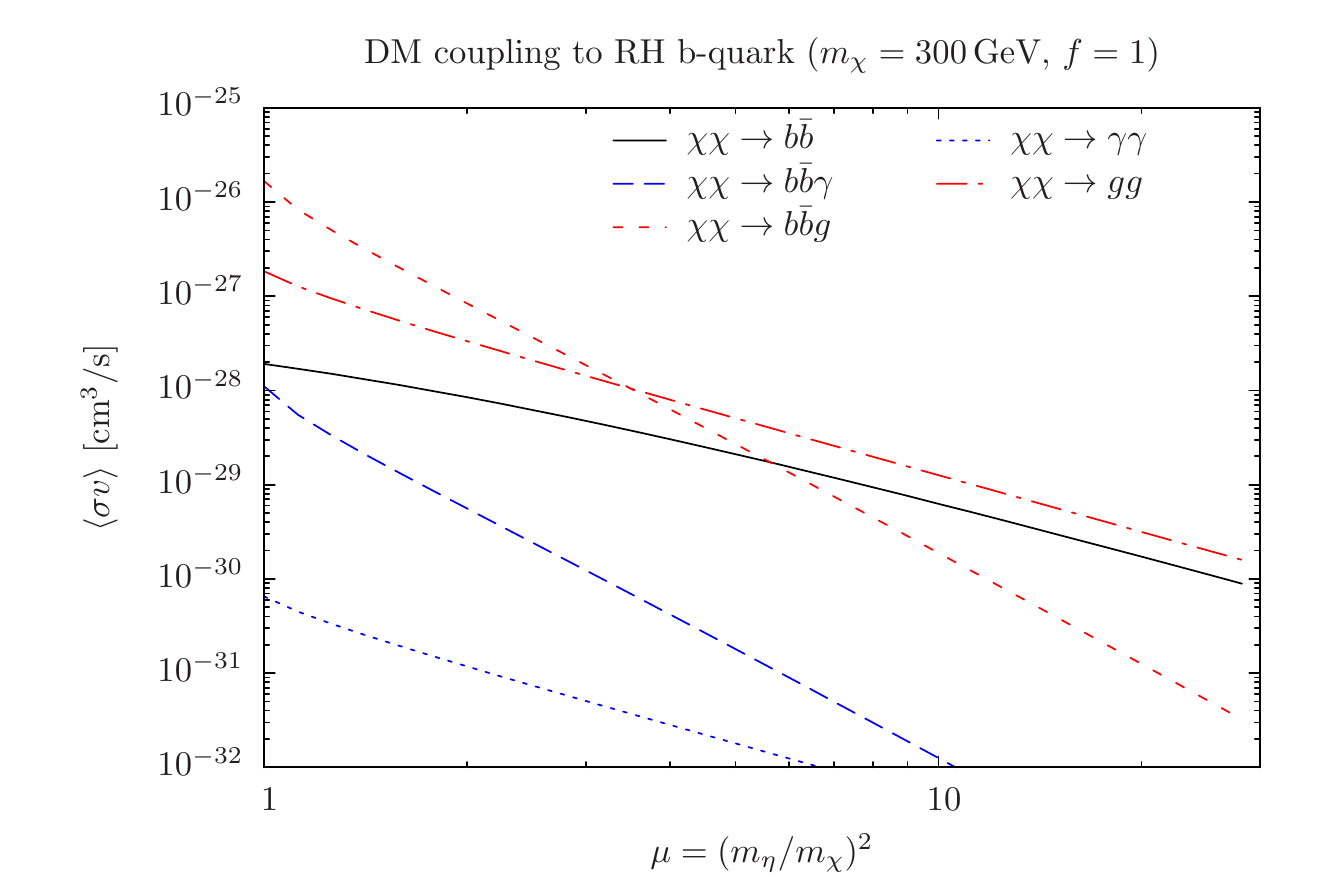}
\caption{Annihilation cross sections of a dark matter particle which couples via a Yukawa interaction to the right-handed bottom-quarks and a coloured scalar $\eta$. The left plot shows the cross sections as a function of the dark matter mass for fixed mass ratio squared $\mu=1.1$, while the right plot as a function of the mass ratio squared $\mu$ for fixed dark matter mass $m_\chi=300$ GeV. In both cases we adopted for definiteness a coupling $f=1$.}
\label{fig:coupling}
\end{figure*}

\par The relative importance of the different annihilation channels strongly  depends on the masses of the particles involved in the annihilation, i.e.~$m_{\chi}, \, m_{\eta}$ and $m_{\psi}$. We show in Fig.~\ref{fig:coupling} the dependence of the different cross sections on the dark matter mass $m_{\chi}$ (left plot) and the degeneracy parameter $\mu$ (right plot) for the case of couplings to $b$-quarks. 
 It should be noted that the gamma-ray spectrum near the 
endpoint at $m_\chi$ is dominated by hard photons from internal 
bremsstrahlung or by monochromatic line photons even when the 
corresponding cross sections are one to two orders of magnitude below 
the total cross section \cite{Bringmann:2012vr,Bringmann:2012ez}. All the annihilation channels, with the exception of $b \bar b$, have cross sections scaling as $1/m_{\chi}^2$, while the annihilation into  $b \bar b$ scales as $1/m_{\chi}^4$ (assuming that the annihilation is dominated by the s-wave contribution, otherwise it is also suppressed by $1/m_{\chi}^2$). Furthermore, replacing the photons with  gluons in the $b \bar b \gamma$ or the $\gamma \gamma$ final state changes the cross section just by a constant,  due to the different colour factors and to the different values of the gauge coupling constants (note that the slightly different dependence on the mass visible in Fig.\,\ref{fig:coupling} is due to the running of $\alpha_s$). 

On the other hand, the two-body cross sections scale as $1/\mu^2$ in the limit of large $\mu$, while the three-body processes scale as $1/\mu^4$. Therefore, the $2\rightarrow 3$ annihilations get enhanced when $\mu$ is small, whereas the loop annihilations into two gauge bosons, $\chi\chi\rightarrow \gamma\gamma, gg$, become more important and dominate as $\mu$ increases.  In particular,  photons produced in the process $\chi \chi \rightarrow \gamma \gamma $ outshine the photons from the $b \bar b \gamma$ final state for $\mu \gtrsim 25$, while internal bremsstrahlung  is dominant at lower values of $\mu$. In order to take this behaviour into account and present meaningful limits on the whole parameter space we will always include the full high-energy spectrum from the line and the internal bremsstrahlung in our analysis. For a more detailed discussion of the dependence of the $2\rightarrow 3$ annihilation processes with the different masses of the model, see \cite{Garny:2011ii}.

\par In our discussion we have concentrated on a toy model consisting of a Majorana dark matter particle which couples to a fermion and a scalar. However, we note that this toy model can be realised in concrete particle physics scenarios, for example in the coannihilation region of the Minimal Supersymmetric Standard Model, or in certain scenarios with right-handed neutrino dark matter \cite{Bergstroem2012}.

\subsection{Thermal freeze-out}

\par One of the strongest arguments in favour of WIMP dark matter is that these particles can be thermally produced in the early universe and generate after freeze-out a relic population with an abundance which is naturally close to the cold dark matter abundance measured by Planck, $\Omega h^2=0.1199\pm 0.0027$ \cite{PlanckCollaboration2013}. Since the parameter space of interest for this paper consists in the region where the dark matter particle $\chi$ and the scalar $\eta$ have fairly degenerate masses, coannihilations can be relevant and the thermal freeze-out must be treated with special care \cite{Griest:1990kh}. In order to take all the relevant processes into account we have used micrOMEGAs2.4 \cite{Belanger:2010gh} to calculate the relic density in a fully numerical way. This approach has been checked against calculations performed with a simplified semi-analytic treatment \cite{Ellis:2001nx} (see \cite{Garny:2012eb} for a detailed description of the calculation of the relic density).

\par The parameter space of the model is spanned by the coupling constant, $f$, the dark matter mass, $m_\chi$, and the scalar mass, $m_\eta$. Then, requiring the observed relic density fixes the coupling constant $f=f_{th}(m_\chi,m_\eta)$, which leaves a parameter space spanned by $m_\chi$ and $m_\eta$ or, alternatively, by $m_\chi$ and $\mu\equiv m^2_\eta/m^2_\chi$. Nevertheless, not all this parameter space is theoretically accessible: coannihilations become more and more important as the mass splitting is reduced and therefore the measured dark matter relic density can only be obtained for smaller and smaller values of the coupling constant $f$.  For very degenerate scenarios, namely $m_{\eta}/m_{\chi} \lesssim 1.1$, we find that coannihilations become so efficient that the total dark matter relic density can no longer be produced via thermal freeze-out unless $m_{\chi} \gtrsim 200$ GeV (50 GeV) when the dark matter couples to quarks (leptons).
This interesting  feature, the  absence of light relics, can  be understood  in terms of a simplified argument. In the coannihilation regime, the effective thermal cross section is given by
\begin{equation}
\sigma v_{eff}=\sigma v(\chi \chi) + \sigma v (\chi \eta) e^{-\frac{m_\eta - m_{\chi}}{T} }+ \sigma v (\eta \eta) e^{-\frac{2 (m_\eta - m_{\chi})}{T}}\;,
\end{equation}  
where processes including the coannihilating particle $\eta$ are Boltzmann-suppressed.  In order to get a clearer idea of the mass and coupling dependence, we can isolate these and express the $\chi \chi$ cross section as $\sigma v(\chi \chi) = \frac{f^4}{m_{\chi}^2} C_{\chi \chi}$ (and likewise for the other processes), where $C_{\chi \chi}$ is a function of mass and temperature ratios only. Absorbing the Boltzmann factors into the $C_{ij}$ functions and using the relation between the relic abundance and the effective cross section we obtain
\begin{equation}
\Omega h^2 \sim \frac{1}{\sigma v_{eff}}=\frac{m_{\chi}^2}{f^4 \, C_{\chi \chi} + f^2 \,g^2\, C_{\chi \eta} + g^4\, C_{\eta \eta}}\;,
\label{eqn:Omega}
\end{equation} 
where $g$ denotes a gauge coupling. Inspecting this equation, it becomes obvious that, unless~$g^4 C_{\eta \eta}=0$,  for arbitrary small masses the relic abundance will always drop below the observed value even if $f=0$ . As in our scenario, with only a mild Boltzmann-suppression and strong coupling, the term independent of $f$ is rather large, the lower bound  can be $\mathcal{O}(100 \text{ GeV})$. Naturally, this simplified argument only captures the qualitative behaviour of our full calculation, and it should be noted that in our numerical calculations we assume $f\gtrsim 10^{-4}$ to ensure that conversions among $\chi$ and $\eta$ remain in equilibrium during freeze-out. Nevertheless, it follows then that mass-degenerate scenarios with dark matter particles with masses violating the lower bound discussed above must include additional production mechanisms apart from thermal production in order to reproduce the cold dark matter abundance inferred from observations.

\section{Experimental searches}\label{searches}

\par Within the class of models discussed above, the most prominent signature is the hard gamma-ray feature arising from internal bremsstrahlung in the annihilation $\chi\chi\to q\bar q \gamma$ (or $\chi\chi\to \ell\bar \ell \gamma$). The gamma-ray spectrum is sharply peaked close to the cutoff energy given by the dark matter mass, and leads to a signature that is basically indistinguishable from a gamma-ray line with present instruments such as Fermi-LAT or H.E.S.S.. On the other hand, the light mediator particle $\eta$ has also a strong influence on possible signals at direct detection experiments as well as colliders. In the following, we briefly discuss the various experimental probes and describe some details of the analysis. While the analysis of spectral features from internal bremsstrahlung is essentially insensitive to the nature of the fermion $\psi$ produced in association with the photon (as long as $m_\psi\ll m_\chi$), the complementary probes can depend strongly on the fermion type. As representative examples we focus mainly on $\psi=u, b$, and briefly discuss the leptonic case $\psi=\mu$ later on.

\subsection{Spectral feature from internal bremsstrahlung}
\label{sec:feature}

\par The excellent energy and angular resolutions of Fermi-LAT enable the search for spectral features in the gamma-ray spectrum. Following this strategy, 43 months of Fermi-LAT data were analysed in \cite{Bringmann:2012vr} to search for spectral features from internal bremsstrahlung coming from directions in the sky with optimal expected signal-to-background ratio in the range $40-300$ GeV (see also \cite{Weniger:2012tx,Fermi-LAT:2013uma} for similar analyses aiming at monochromatic gamma-rays). Recently, also the H.E.S.S.~collaboration has analysed 112h of data from the central galactic halo collected during four years \cite{Abramowski:2013ax}. The search region consists of a narrow cone with $1^\circ$ radius around the galactic centre, but excluding the galactic plane $|b|<0.3^\circ$ in order to minimise possible astrophysical backgrounds. The main background is then expected to consist of misidentified cosmic-ray protons, which were suppressed by applying suitable cuts as far as possible. The residual observed flux was then used to search for spectral features and place upper limits on annihilation cross sections for monochromatic gamma rays as well as internal bremsstrahlung in the range $500$ GeV--$25$ TeV.

\par The existing limits can be improved by future observations with a better energy resolution and/or by increasing the statistics via instruments with large effective observation area \cite{Bergstrom:2012vd}. For the case of imaging air Cherenkov telescopes, also an improvement in the proton/photon discrimination can help to further suppress the background flux. As two important examples, we will consider the planned GAMMA-400 \cite{Galper:2012fp} satellite mission as well as the Cherenkov Telescope Array (CTA) \cite{Consortium:2010bc}. While the former is designed to achieve an energy resolution at the percent level, the latter mainly profits from the huge effective area, $A_{eff}\simeq 0.02, 0.3, 2.3$ km$^2$ at 100 GeV, 1 TeV and 10 TeV energies, respectively \cite{Bernlohr:2012we}. For comparison, the effective area of GAMMA-400 will be ${\cal O}(1$ m$^2)$ and the energy resolution of CTA is estimated as $\sigma/E\sim 25\%, 10\%, 5\%$, respectively, at the same energies as given above.

\medskip

\par For internal bremsstrahlung, the primary gamma-ray spectrum is peaked at energies $E_\gamma^{max}$ slightly below the dark matter mass $m_\chi$, and has an intrinsic width which depends on the mass ratio $m_\eta/m_\chi$. For example, $x_{max}=0.95^{+0.04}_{-0.19}, 0.89^{+0.09}_{-0.22}, 0.79^{+0.16}_{-0.27}$ for $m_\eta/m_\chi=1.01, 1.1$ and 2, respectively. Here, $x\equiv E_\gamma/m_\chi$ and the upper and lower values denote the half-width at half maximum above and below the peak of the spectrum. In addition, as discussed above, for $2\lesssim m_\eta/m_\chi\lesssim 5$ the spectrum consists of a superposition of internal bremsstrahlung and a monochromatic line at $x=1$ with comparable contributions. 

\par In order to derive appropriate limits for the specific shape of the spectrum in each case, we compute the differential gamma-ray flux arriving at the Earth from an angle $\xi$ with respect to the Galactic centre,
\begin{equation}
  \frac{d \Phi}{dE d\Omega } = \frac{1}{4\pi} \left( \frac{d\sigma v_{q\bar q\gamma}}{dE} + 2\sigma v_{\gamma\gamma}\delta(E-m_\chi) \right) \, \int_0^\infty ds \, \frac12 \left(\frac{\rho_{dm}(r)}{m_\chi}\right)^2 \,,
\end{equation}
where $r=\sqrt{(r_0-s \cos\xi)^2+(s\sin\xi)^2}$, $r_0=8.5$\,kpc. For definiteness, we assume a radial dark matter distribution given by the Einasto profile
\begin{equation}
  \rho_{dm}(r) \propto \exp \left(-\frac{2}{\alpha_E} \left(\frac{r}{r_s}\right)^{\alpha_E} \right)%\,,
\end{equation}
with $\alpha_E=0.17$ and scale radius $r_s=20$\,kpc \cite{Diemand:2008in}, normalised to $\rho_{dm}(r_0)=0.4$\,GeV$/$cm$^3$. Next, we convolute the primary spectrum for $40$ GeV $ \leq m_\chi \leq 10$ TeV and $1.01 \leq m_\eta/m_\chi \leq 10$ with the expected energy resolution for Fermi-LAT \cite{Ackermann:2012kna} and H.E.S.S.~\cite{Abramowski:2013ax}, respectively. We then perform a binned profile likelihood analysis to set one-sided 95\% confidence level (C.L.) upper limits on a signal contribution from dark matter on top of a smoothly varying background spectrum fitted to the Fermi-LAT data taken from \cite{Weniger:2012tx} (search region 3, Pass7 SOURCE sample) and the H.E.S.S.~data from \cite{Abramowski:2013ax} (CGH region). The background parametrisation and choice of the energy range included in the fit is chosen as in \cite{Bringmann:2012vr} for Fermi-LAT and \cite{Abramowski:2013ax} for H.E.S.S., and we refer to these references for details on the statistical analysis. We note that, for a pure monochromatic gamma-ray line, we checked that our limits agree at the 20\% level with those from \cite{Weniger:2012tx,Abramowski:2013ax} for all energies. We also note that the line limits presented recently by the Fermi-LAT collaboration \cite{Fermi-LAT:2013uma} are comparable to those derived in \cite{Weniger:2012tx} for the Einasto profile.

\par We also derive expected limits by generating a large number of data samples under the background-only hypothesis, with Poissonian fluctuations around the counts in each bin corresponding to the assumed background flux. We then derive upper limits on the signal contribution for each sample and obtain the expected limits by taking the average on a logarithmic scale. For the case of GAMMA-400, we adopt the search region, observation time, background flux, effective area and energy resolution from \cite{Bergstrom:2012vd}, but use the energy window from \cite{Bringmann:2012vr} which is more suitable for internal bremsstrahlung. We again checked agreement with \cite{Bergstrom:2012vd} for the case of gamma-ray lines. To obtain expected limits for CTA, we use the projected effective area and energy resolution presented in \cite{Bernlohr:2012we} (MPIK settings), and assume 50h observations of a region with a $2^\circ$ radius around the galactic centre. For the background we take into account the gamma-ray emission close to the galactic centre measured by H.E.S.S.~as well as the cosmic-ray electron and proton fluxes \cite{Bringmann:2011ye}. For the latter we assume a proton/photon discrimination efficiency of 1\%, and furthermore assume that 80\% of all incoming events are observed. We take as background a power law with free slope and normalisation, which are treated as nuisance parameters, and use a sliding energy window $(m_\chi/\epsilon^{0.7},m_\chi\epsilon^{0.3})$ with $\epsilon=3$ $(4)$ at $m_\chi=1$ $(10)$ TeV. See also \cite{Aleksic:2012cp} for a related approach.

\subsection{Secondary gamma rays}

\par A peculiar feature of internal bremsstrahlung is that it produces a sharp spectral feature in the gamma-ray spectrum together with a comparably small flux of continuum gamma rays. The latter originate from the decay and fragmentation of the SM particles produced in all annihilation processes. Typically, this continuum component overwhelms the monochromatic flux from $\chi\chi\to\gamma\gamma$ because the latter channel is suppressed by a loop factor and two powers of the fine-structure constant. In contrast, the cross section for  $\chi\chi\to q\bar q\gamma$ depends linearly on $\alpha_{em}$. Furthermore, since the channel $\chi\chi\to q\bar q$ is helicity-suppressed, the continuum gamma-ray spectrum arises predominantly from the annihilation channel $\chi\chi\to q\bar q g$ for the most interesting part of the parameter space ($m_q\ll m_\chi$ and $m_\eta\lesssim 2 m_\chi)$. The ratio of cross sections is given by \cite{Garny:2011ii}
\begin{equation}\label{eq:ratioIB}
  \frac{\sigma v(\chi\chi\to q\bar q\gamma)}{\sigma v(\chi\chi\to q\bar q g)} = \frac{Q_q^2\alpha_{em}}{C_F\alpha_s} \simeq 3\%(0.7\%)\,,
\end{equation}
where $C_F=4/3$ and the numerical values correspond to up(down)-type quarks, using $\alpha_s$  at $\mu=300$ GeV. For somewhat larger splittings $m_\eta\gtrsim 2 m_\chi$, the annihilation channel $\chi\chi\to gg$ contributes significantly to the secondary gamma-ray spectrum (cf.~Fig.~\ref{fig:coupling}). In the limit $m_\eta\gg  m_\chi \gg m_q$, this channel even becomes dominant, due to the helicity and velocity suppression of the tree level annihilation. In this regime, the relative strength of the hard spectral feature to the continuous secondary flux is determined by the ratio
\begin{equation}\label{eq:ratioLine}
  \frac{2\,\sigma v(\chi\chi\to \gamma\gamma)}{\sigma v(\chi\chi\to g g)} = \frac{Q_q^4N_c^2\alpha_{em}^2}{\alpha_s^2} \simeq 1\%(0.06\%)\,,
\end{equation}
for up(down)-type quarks, using $\alpha_s$  at $\mu=300$ GeV.

\par To obtain constraints on the secondary gamma-ray flux from $q\bar q g$ (and $q\bar q$, $gg$) we use Fermi-LAT observations of dwarf galaxies \cite{GeringerSameth:2011iw,Ackermann:2011wa}, $\sigma v<5\cdot 10^{-30}$ cm$^3/$s$/$GeV$^2 \times 8\pi m_\chi^2/N_\gamma$ where $N_\gamma$ is the number of photons per annihilation in the range $1-100$ GeV (see also \cite{Bringmann:2012vr}). Note that the same annihilation channels also yield a flux of cosmic antiprotons. Corresponding limits obtained from the antiproton-to-proton fraction measured by PAMELA \cite{Adriani:2010rc}  typically turn out to be  slightly weaker \cite{Garny:2012eb} than the dwarf limits but can be more stringent for very small mass splittings \cite{Asano:2011ik}, depending on the relatively large uncertainties in the propagation of charged cosmic rays in the galactic magnetic field. Furthermore, the comparisons of limits from dwarf galaxies and the Milky Way should always be taken with a grain of salt as there could be additional independent boost factors for both systems.

\subsection{Direct searches}

\par As noticed originally in Ref.~\cite{Hisano:2011um} and further pursued in our previous works \cite{Garny:2012eb,Garny:2012it}, the existence of a scalar particle $\eta$ just slightly heavier than the dark matter particle $\chi$ leads to dramatic enhancements in both spin-dependent and spin-independent cross-sections. This has important consequences for the constraints on mass-degenerate dark matter scenarios coming from direct searches. Here, we follow the exact same procedure as outlined in \cite{Garny:2012eb} for the computation of recoil rates (including spin-dependent and spin-independent contributions) and the derivation of bounds on the Yukawa coupling $f$ defined in Sec.~\ref{ppmodel}. The direct detection limits throughout this work refer to the latest XENON100 data \cite{XENON100_2012} only, shown in \cite{Garny:2012eb} to be the most stringent at present across the whole parameter space of mass-degenerate scenarios. For prospects we use the expected sensitivity of XENON1T. Lastly, all limits shown are the most conservative within the band of nuclear and astrophysical uncertainties considered in the reference above.

\subsection{Collider searches}

\par Within the model considered here, the interactions of the dark matter particle $\chi$ with the SM fermions are mediated by the scalar particle $\eta$. Since we are most interested in the case where $m_\eta$ is comparable to $m_\chi$, the scalar $\eta$ may be considered as a light mediator. Therefore, the collider phenomenology is very different from the effective theory approach that would be applicable for $m_\eta\gg m_\chi$ \cite{Rajaraman:2011wf,Fox:2011pm}. Instead, the scalar $\eta$ could be pair-produced directly in proton-proton collisions. For the cases where the light mediator couples $\chi$ to quarks, it carries a colour quantum number and can therefore be copiously produced e.g.~via $gg\to \eta\bar\eta$. The subsequent decay $\eta\to\chi q$ then leads to typical signatures with two jets and missing transverse energy. If, on the other hand, the mediator couples $\chi$ to leptons, pair production via the Drell-Yan process leads to much smaller production cross sections at the Large Hadron Collider (LHC). The signature in this case consists of a pair of oppositely charged leptons and missing transverse energy. The phenomenology thus closely resembles simplified supersymmetric models containing squarks and neutralino, or sleptons and neutralino, respectively, with all other SUSY particles decoupled. 

\par An important difference compared to the supersymmetric case is the production cross section. First, we consider only a single scalar $\eta$ while the simplified supersymmetric models typically contain eight mass-degenerate squarks (up/down, left/right, 1st/2nd generation, see \cite{Mahbubani:2012qq} for a discussion of non-degenerate squarks) or two sleptons (1st/2nd generation). Second, apart from the production diagrams involving only gauge interactions, which are identical to the SUSY case, there exist also processes such as $qq\to \eta\eta$ that are mediated by $\chi$ in the t-channel and the cross section of which is proportional to $f^4$. In the SUSY case, this coupling is fixed by the hypercharge, e.g.~$f_{SUSY}\simeq 0.33,0.16,0.5$ for the case where dark matter couples to right-handed up-type quarks, down-type quarks or charged leptons, respectively. In this work we consider the coupling $f$ as a free parameter of the toy model, and find that e.g.~the process $uu\to \eta\eta$ can give a sizable contribution to the production cross section in proton-proton collisions when $f\gtrsim 0.5-1$. One possibility to fix the coupling $f$ is to require that, for a given set of masses $m_\chi$ and $m_\eta$, thermal freeze-out produces a relic density in agreement with cosmic microwave background observations, i.e.~$f=f_{th}(m_\chi,m_\eta)$. When taking coannihilations into account we typically find $f_{th}< f_{SUSY}$ for small splittings $m_\eta/m_\chi\lesssim 1.2$ and $m_\chi\lesssim 1$ TeV. On the other hand, for larger masses and/or mass splittings the coupling required for thermal production can be of order one or larger. 

\par To obtain an estimate for the regions of parameter space excluded by LHC searches we consider, for the case of a coloured scalar, the $\alpha_T$-analysis \cite{Chatrchyan:2013lya} by the CMS collaboration based on 11.7 fb$^{-1}$ at 8 TeV energy. This analysis is designed to search for hadronic final states with missing transverse energy. We use the upper limit on the production cross section obtained for a simplified model containing squarks and neutralino, and derive an exclusion limit by requiring that the production cross section computed for a single species of coloured scalars taken from \cite{susyxsectionwg} lies below the upper limit. We also take the effect of the coupling $f_{th}(m_\chi,m_\eta)$ into account in an approximate way by computing the corresponding contribution to the production cross section at leading order using CalcHEP \cite{Pukhov:1999gg,Pukhov:2004ca}. We caution the reader that the exclusion limits obtained in this way should be regarded as an estimate. A dedicated analysis would certainly be desirable, but goes beyond the scope of this work. 

\par For small splittings between the dark matter particle and the scalar $\eta$, the jets from $\eta\to\chi q$ become so soft that the sensitivity is greatly reduced. This effect can be partly compensated by taking into account recoil against hard jets emitted by one (or several) of the coloured particles (e.g.~initial state radiation). The latter will be included in the upcoming LHC analyses for SUSY models. Since a dedicated analysis for the model considered here is beyond the scope of this work, we adopt the corresponding limits obtained in \cite{Dreiner:2012gx} from  $7$ TeV data for a single species of coloured scalars decaying into light quarks and missing energy. When the masses of $\eta$ and $\chi$ are quasi-degenerate ($m_\eta-m_\chi\lesssim 20$ GeV), even monojet searches can become relevant, since the quarks produced in the decay $\eta\to\chi q$ have typically very small energies \cite{Dreiner:2012gx}.

\par For the case where dark matter couples to right-handed muons, we use the LEP limits from \cite{lepslepton} as well as the ATLAS search \cite{atlasslepton} for direct production of right-sleptons to obtain approximate collider limits.

\section{Results}\label{results}

\par In this section we compare the limits obtained from the various experimental probes described above and discuss their interplay at present, as well as in the future. We mainly focus on the case where the scalar $\eta$ mediates an interaction of dark matter with quarks, and then briefly comment on the main differences in the leptonic case.

\subsection{Coupling to quarks}

\begin{figure*}[htpb]
%\centering
\includegraphics[width=0.9\textwidth]{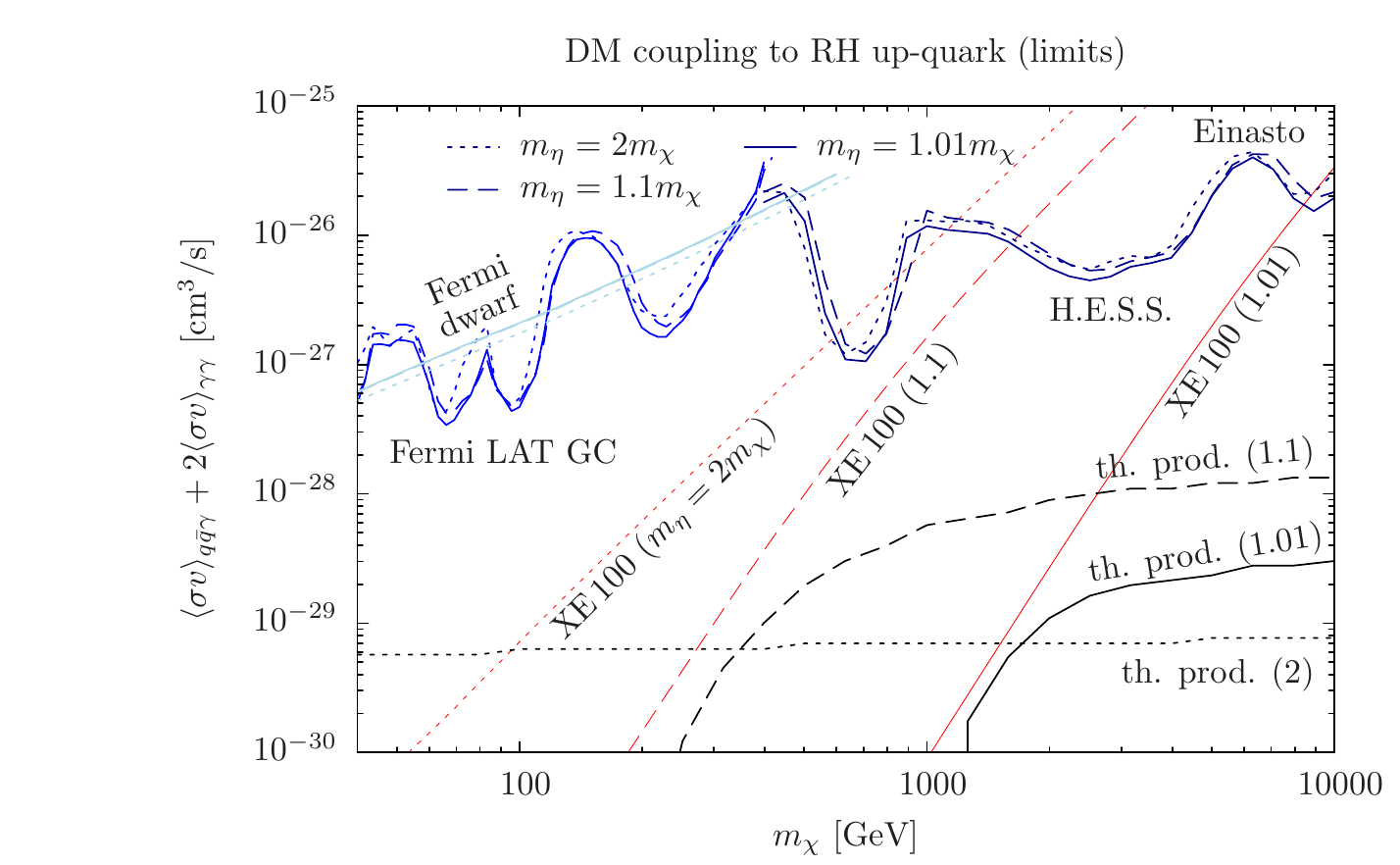}
\\[2ex]
\includegraphics[width=0.9\textwidth]{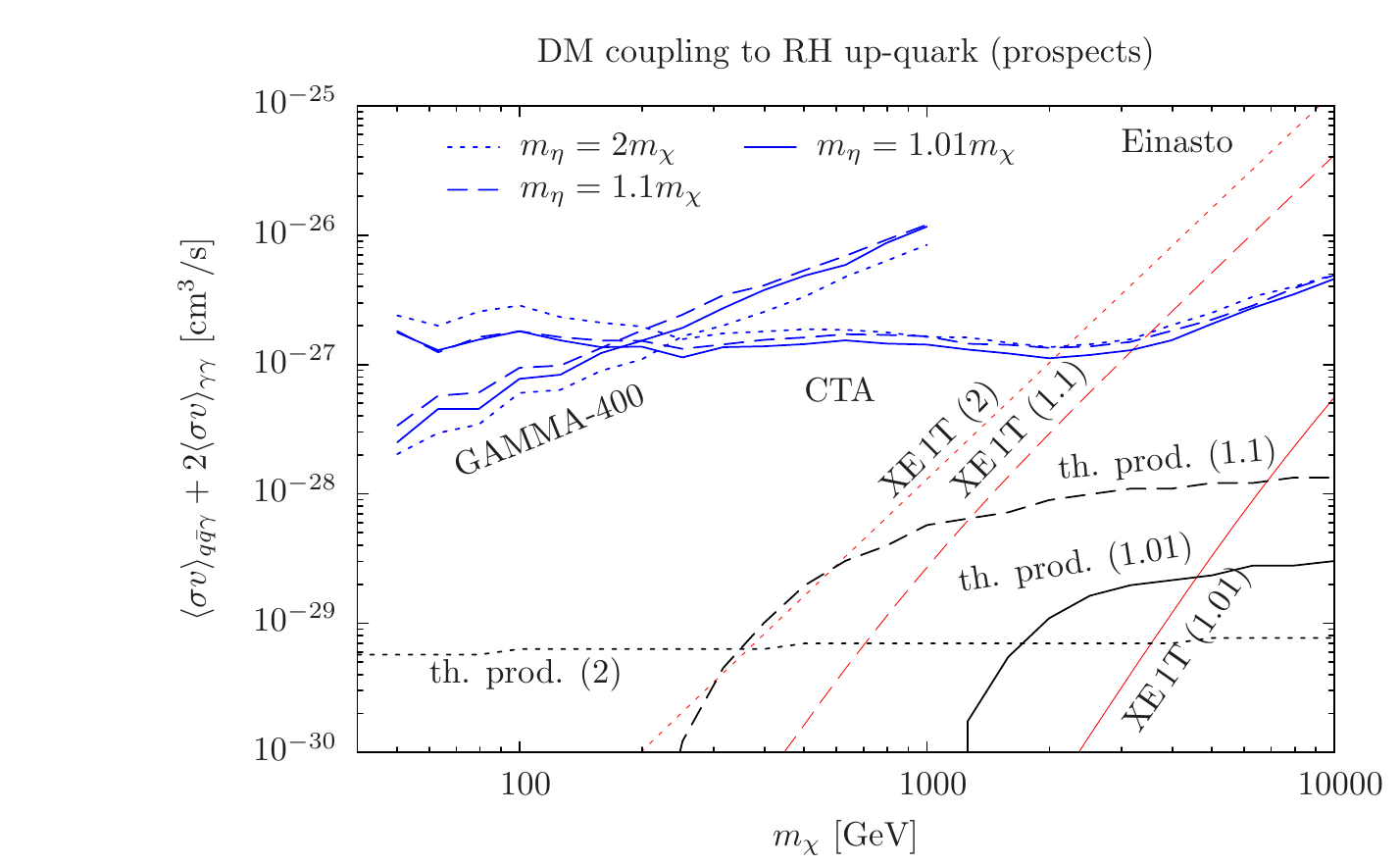}
\caption{Upper limits ($95\%$ C.L.) on the annihilation cross section obtained from searches for a spectral feature in the gamma-ray spectrum using Fermi-LAT and H.E.S.S.~observations (blue lines), for three values of the mass splitting $m_\eta/m_\chi=1.01,1.1,2$ and assuming an Einasto profile. Also shown are complementary constraints from direct searches (XENON100, red lines) and constraints on secondary gamma rays from Fermi-LAT observations of dwarf galaxies (light blue lines), both of which were translated into limits on the annihilation cross section $\langle\sigma v_{q\bar q\gamma}\rangle+2\langle\sigma v_{\gamma\gamma}\rangle$. The black lines indicate the cross section expected for a thermal relic. The lower frame shows an estimate for the upper limit that can be achieved by searches for spectral features by GAMMA-400 and CTA, respectively, as well as the prospect for XENON1T. Here we assumed that dark matter interacts with right-handed up-quarks.} 
\label{fig:IB}
\end{figure*}

\begin{figure*}[htp]
%\centering
\includegraphics[width=0.9\textwidth]{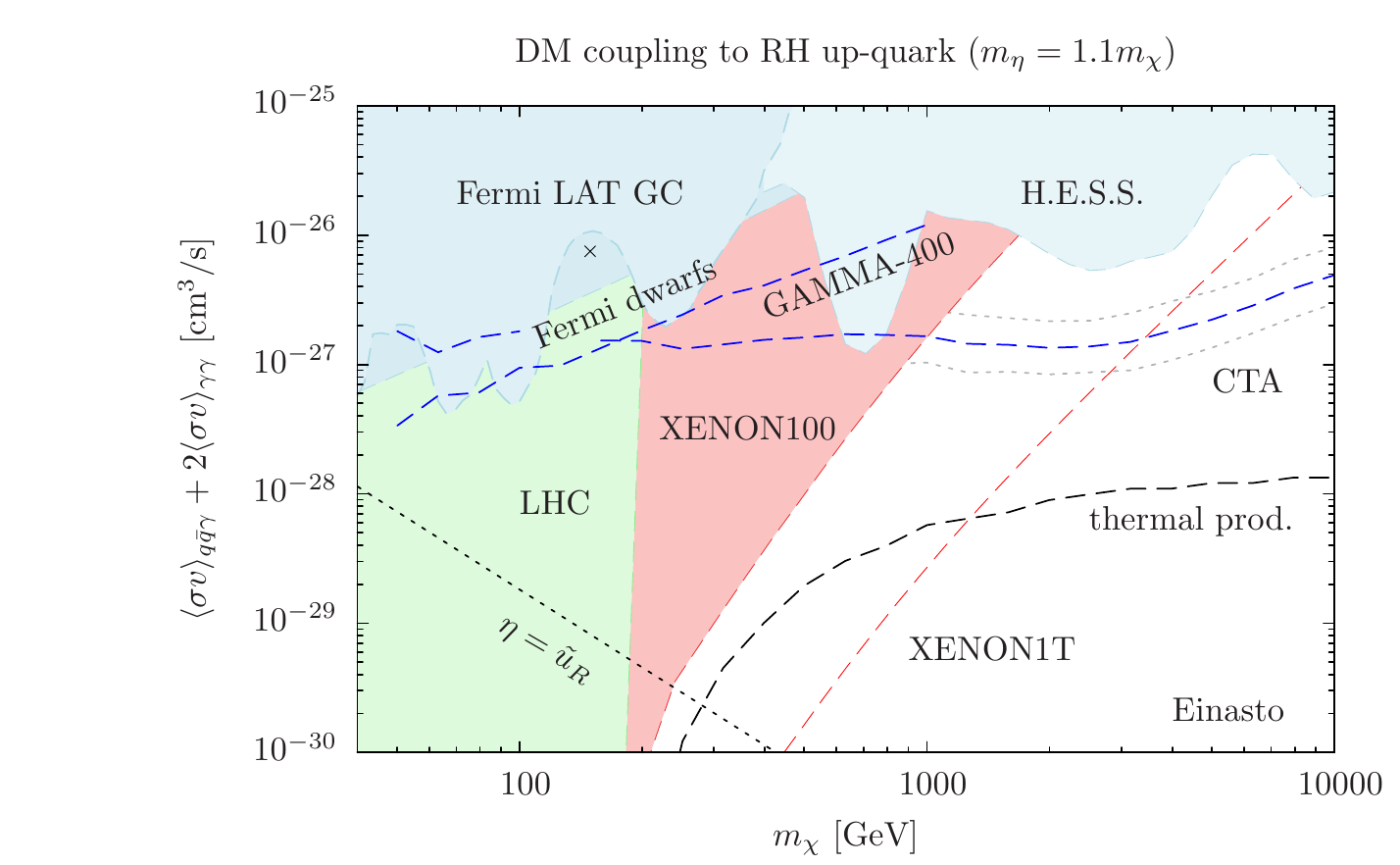}
\\[2ex]
\includegraphics[width=0.9\textwidth]{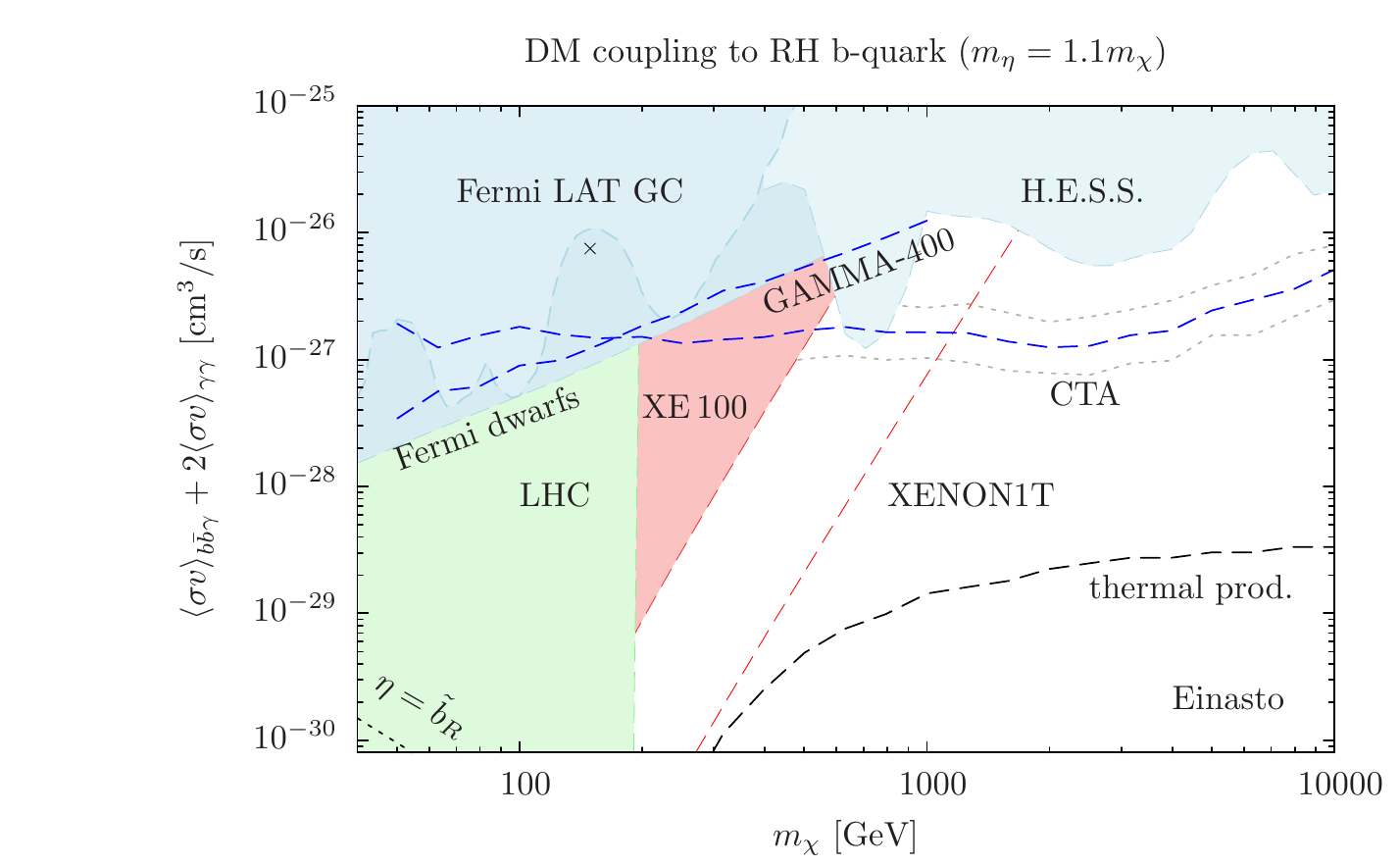}
\caption{Comparison of future prospects for searches for spectral features in gamma-rays (GAMMA-400, CTA) and direct searches (XENON1T) in the case where dark matter couples to right-handed up- and bottom-quarks, respectively in the top and bottom frames, and $m_\eta/m_\chi=1.1$. The shaded areas are excluded by various present experiments as discussed in the text. The dashed black line shows the cross section expected for a thermal relic, and the dotted black line refers to the case where the Yukawa coupling is fixed to $f=0.33$ $(0.16)$ for $u$ ($b$), as is appropriate in the supersymmetric case. The dotted grey lines correspond to the $\pm1\sigma$ range for the expected upper limit for CTA.} 
\label{fig:prospects}
\end{figure*}

\subsubsection{Current limits}

\par The search for characteristic features in the gamma-ray spectrum from near the galactic centre is directly sensitive to the annihilation cross section of $\chi\chi\to q\bar q\gamma$ responsible for internal bremsstrahlung. If $m_\eta\gtrsim 2m_\chi$, also the annihilation cross section into gamma-ray lines gives a non-negligible contribution, and we therefore show in the following upper limits on their sum, $\sigma v_{q\bar q\gamma}+2\sigma v_{\gamma\gamma}$. For the same reason we also adapted the previous analyses of Fermi-LAT data \cite{Bringmann:2012vr} and from H.E.S.S.~\cite{Abramowski:2013ax}, that were done assuming a pure internal bremsstrahlung spectrum. In Fig.~\ref{fig:IB} (top), we show the resulting upper limits ($95\%$ C.L.) obtained for $q=u_R$ and three values of the mass ratios $m_\eta/m_\chi=1.01,1.1,2$, ranging from $10^{-27}-10^{-26}$ cm$^3/$s for $m_\chi=40-10^4$ GeV. As was observed previously \cite{Bringmann:2012vr}, the dependence on the splitting is rather mild. When taking only internal bremsstrahlung into account, the limits degrade slightly when increasing $m_\eta/m_\chi$ since the spectrum becomes less sharply peaked \cite{Bringmann:2012vr}. However, adding the line contribution counteracts this tendency because it sharpens the gamma-ray feature, and it becomes more important when increasing $m_\eta/m_\chi$ (see Fig.~\ref{fig:coupling}). The superposition of these effects changes the dependence of the limits on $m_\eta/m_\chi$. They degrade slightly up to a certain value $m_\eta/m_\chi\sim {\cal O}(2-4)$, and then become slightly stronger again. Typically, this reduces the dependence on the mass ratio even further (cf.~top frame of Fig.~\ref{fig:IB}). As expected, we find that the limits from spectral features also depend only very mildly on the quark flavour as clear from Fig.~\ref{fig:prospects}.  In contrast to the observational limits, the theoretical predictions for a thermal relic or for a fixed coupling depend rather strongly on the mass splitting. 
  
\par
\begin{figure*}[htp]
\centering
\includegraphics[width=0.69\textwidth]{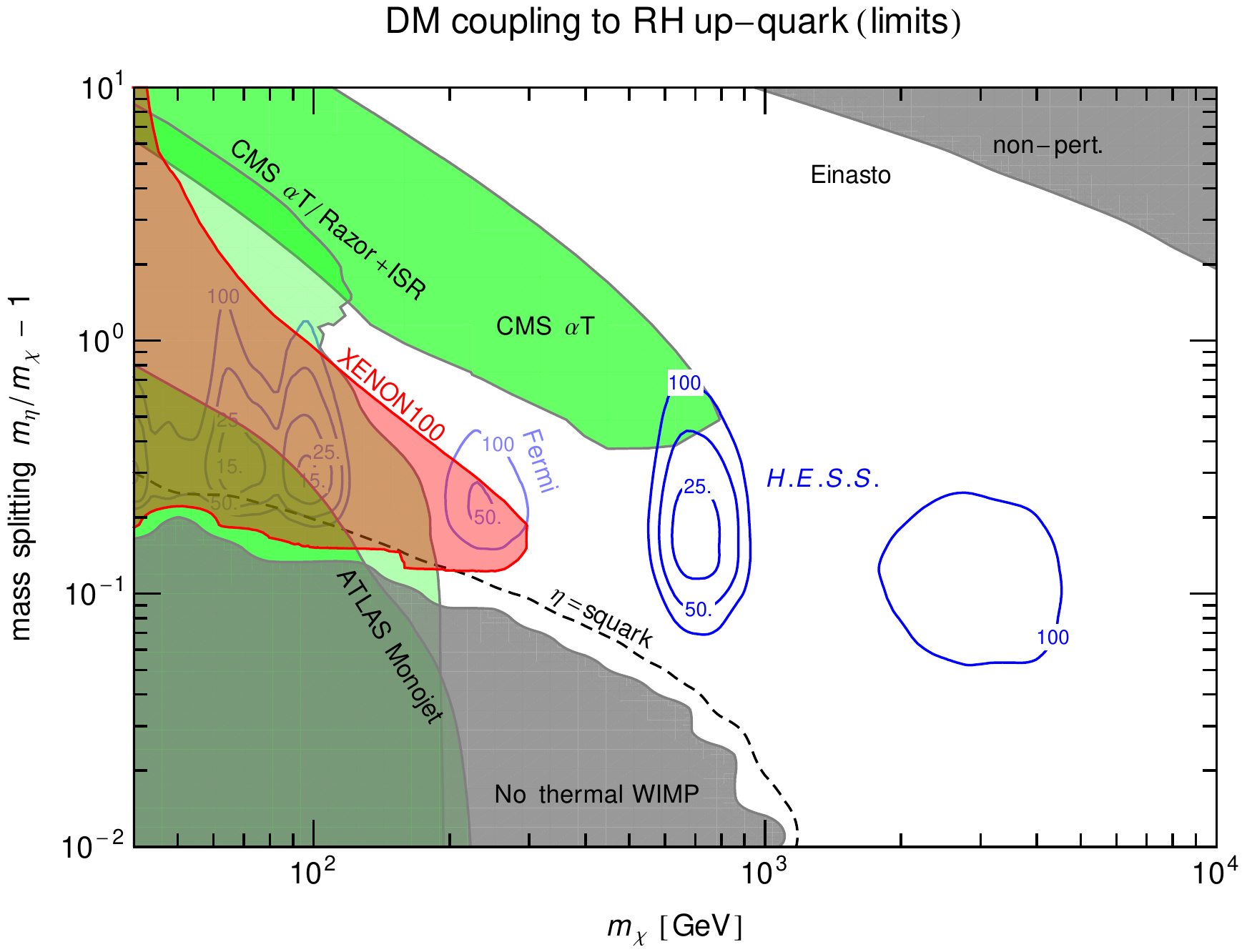}
\\[4ex]
\includegraphics[width=0.69\textwidth]{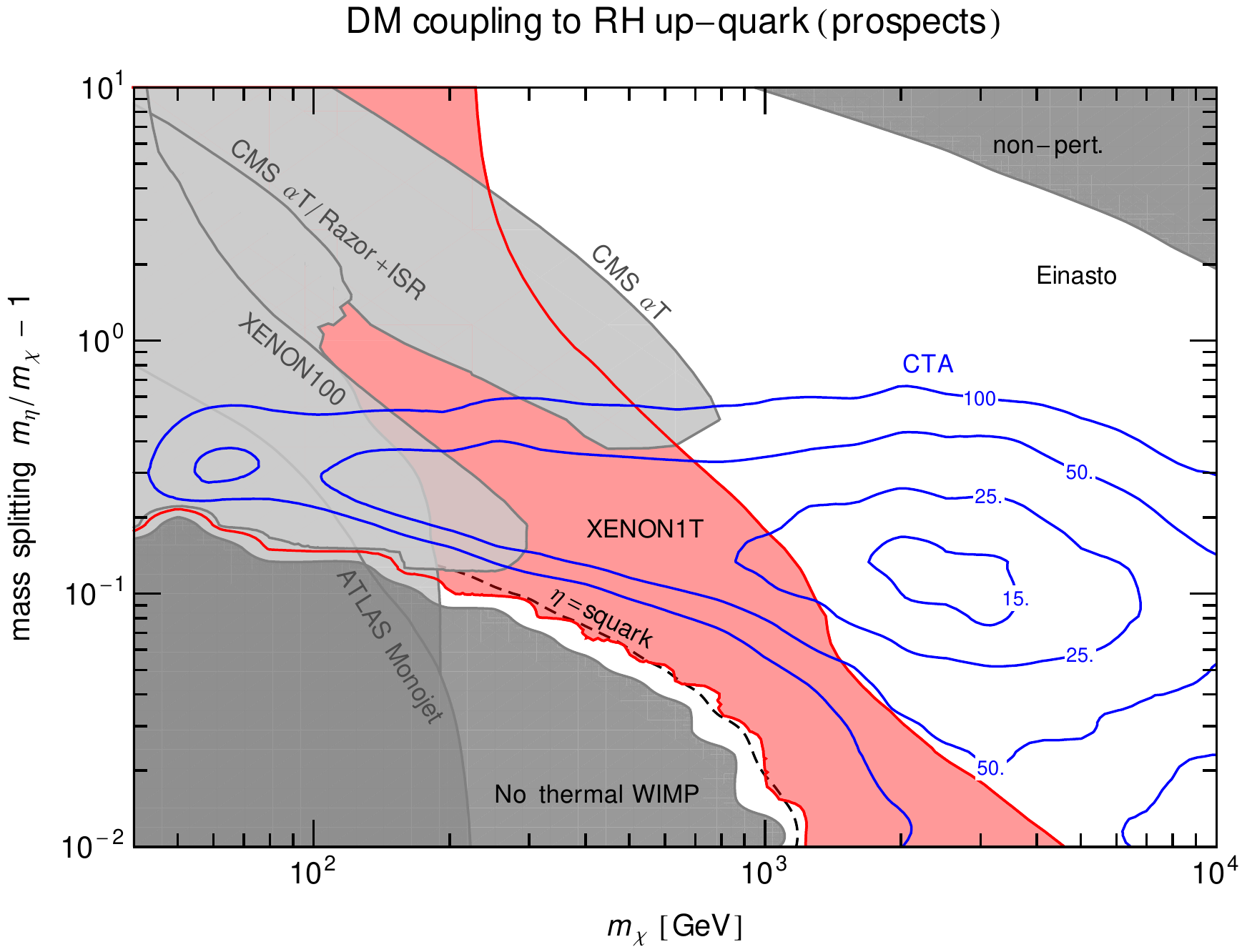}
\caption{Comparison of experimental constraints as a function of the dark matter mass $m_\chi$ and the relative mass splitting $m_\eta/m_\chi-1$. Here we assume that dark matter interacts with right-handed up-quarks and fix the Yukawa coupling $f=f_{th}(m_\eta,m_\chi)$ at each point by requiring that the thermal relic density matches the one derived from the cosmic microwave background. The red region in the upper frame is excluded by XENON100 at $95\%$ C.L., and the green regions indicate exclusions by various collider searches. The contour lines show the ratio $r=\sigma v_{U.L.}/\sigma v_{th}$ of the upper limits on the annihilation cross section obtained from searches for spectral features with Fermi-LAT and H.E.S.S., respectively, and the cross section expected for a thermal relic. The regions inside the contours are excluded if the annihilation signal is boosted relative to the Einasto profile by a factor $BF\geq r$. Within the dark grey regions, thermal freeze-out cannot account for the cold dark matter density (in the lower right corner $\Omega_{th} h^2<0.12$ for all values of the coupling $f$, and in the upper right corner $f$ becomes non-perturbatively large). The dashed black line corresponds to $f_{th}(m_\eta,m_\chi)=0.33(=f_{SUSY})$. The lower frame shows the prospects for the region that can be excluded by XENON1T and the expected limits from CTA.
} 
\label{fig:mDMvsSplittingUR}
\end{figure*}

 It is also important to note that, when the ratio $R\equiv (\sigma v_{q\bar q\gamma}+2\sigma v_{\gamma\gamma})/(\sigma v_{total})$ becomes small enough, the tail of the secondary gamma spectrum starts to leak significantly into the search window for the hard spectral feature. In this case, the search strategy described in section \ref{sec:feature} should not be applied \cite{Bringmann:2011ye}. For $m_\chi \gg m_q$, $R$ is given by Eq.~(\ref{eq:ratioIB}) for $m_\eta/m_\chi\to 1$ and by Eq.~(\ref{eq:ratioLine}) for $m_\eta \gg m_\chi$. We explicitly checked for $m_\chi=100\,$GeV that, when including the tail of the secondary spectrum in the fit, the obtained limits change by less than $30-40\%$ over the range $m_\eta/m_\chi=1.01\dots10$  (corresponding to $R\gtrsim 3\dots 1\%$) for coupling to up-type quarks. For bottom quarks, the correction remains below $50\%$ for $m_\eta/m_\chi\lesssim 1.13$ (corresponding to $R=0.4\%$), but can be larger than a factor two for $m_\eta/m_\chi\gtrsim 1.3$ ($R<0.3\%$), mainly because the line contribution is strongly suppressed because of the smaller charge. For $1\%(10\%)$ splitting, $R>0.5\%$ for $m_\chi>140(200)\,$GeV for coupling to bottom quarks. Note that $R>1(0.4)\%$ within the parameter ranges shown in Figs.~\ref{fig:IB}-\ref{fig:mDMvsSplittingUR} for up(bottom)-quarks.

\par The limits obtained from secondary gamma rays from Fermi-LAT observations of dwarf galaxies are arising mainly from $\chi\chi\to q\bar q g$, and are rather insensitive to $m_\eta/m_\chi$ (cf.~top frame of Fig.~\ref{fig:IB}). The corresponding upper limits on $\sigma v_{q\bar q g}$ are also comparable for $q=u_R$ and $q=b_R$. When translated into limits on $\sigma v_{q\bar q\gamma}$, the dwarf limits  for $q=b_R$ are consequently stronger by a factor of the order of $Q_u^2/Q_d^2=4$ than those for $q=u_R$ (since $\sigma v_{q\bar q\gamma}/\sigma v_{q\bar q g}\sim Q_q^2$). In comparison to the searches for spectral features from the galactic centre region, the dwarf limits are comparable for $q=u_R$, and slightly stronger for $q=b_R$, see Fig.~\ref{fig:prospects}. However, we stress that their relative strength depends on the uncertainties of the dark matter distribution in the galactic centre and in dwarf spheroidal galaxies, respectively.

\par The Yukawa coupling between dark matter and quarks may also be observable via scatterings $\chi N\to \chi N$ off heavy nuclei. Since the mediator $\eta$ appears in the s-channel, the scattering rate is resonantly enhanced for small mass splittings $m_\eta-m_\chi$ and therefore leads to strong limits from direct detection experiments \cite{Hisano:2011um,Garny:2012eb}. For example, for up-quarks, the direct detection limits obtained from XENON100 are stronger than the limits from spectral features for dark matter masses $m_\chi\lesssim 2$ TeV if $m_\eta/m_\chi=1.1$, as shown in the top frame of Fig.~\ref{fig:IB}. For larger splitting, the XENON100 limits are somewhat weaker. On the other hand, for very small splittings ($m_\eta/m_\chi=1.01$) direct detection limits dominate nearly over the whole range we are considering (up to 8 TeV). If dark matter couples to bottom quarks, the limits from direct detection are weaker by about an order of magnitude (cf.~Fig.~\ref{fig:prospects}). Nevertheless, they are still stronger compared to the Fermi-LAT limits on spectral features and on secondary gamma rays, and stronger than (or comparable to) the H.E.S.S.~limits for $m_\chi\lesssim 600$ GeV  if $m_\eta/m_\chi=1.1$. Along the same lines, it is interesting to point out that the tentative spectral feature spotted in Fermi-LAT data at around 130 GeV \cite{Bringmann:2012vr,Weniger:2012tx,Fermi-LAT:2013uma}, indicated by a cross in Fig.~\ref{fig:prospects}, is convincingly excluded by XENON100 (as well as LHC bounds and Fermi-LAT dwarf observations) in hadronic dark matter models.

\par The dependence of the various limits on the dark matter mass and the mass splitting is summarised in Fig.~\ref{fig:mDMvsSplittingUR} (top). In particular, within the region shaded in red the XENON100 bound is stronger than the scattering cross section off Xe nuclei expected for a thermal relic. For the conservative set of nuclear parameters (see \cite{Garny:2012eb}), this region extends up to $m_\chi\lesssim 300$ GeV for a splitting of $15-20\%$. \FloatBarrier Note that, for smaller splittings, the cross sections expected for a thermal relic decrease very quickly due to efficient coannihilations. This overcompensates even the resonant enhancement of the scattering cross section off nuclei for small splittings. In particular, within the grey shaded region coannihilations are so efficient that the relic density falls below the observed dark matter relic density even for tiny Yukawa couplings $f\leq 10^{-3}$. 

\par It also follows from Fig.~\ref{fig:mDMvsSplittingUR} that the sensitivity of gamma-ray observations reaches a maximum for $m_{\eta} / m_{\chi} \approx 1.1$. This optimal value arises from the fact that, for a fixed coupling constant, the annihilation rate into $q \bar q \gamma$ increases as $m_{\eta} / m_{\chi}$ decreases. At the same time, as mentioned before, for decreasing $m_{\eta} / m_{\chi} $  coannihilations become more and more efficient and  a smaller and smaller coupling constant is necessary in order to reproduce the correct relic abundance. Therefore, for a thermally produced dark matter particle, the annihilation cross section into  $q \bar q \gamma$ does not increase monotonically as  $m_{\eta} / m_{\chi}$ decreases, but reaches a maximum, due to the nontrivial dependence of the coupling constant $f_{th}(m_\chi,m_\eta)$ on the masses.

\par In Fig.~\ref{fig:mDMvsSplittingUR} we also show a collection of various collider bounds as described in the previous section. The bounds obtained from a CMS search for hadronic final states and missing transverse energy using the $\alpha_T$ variable \cite{Chatrchyan:2013lya} indicate that dark matter masses up to TeV energies can be excluded when the mass splitting is not too small ($m_\eta-m_\chi\gtrsim {\cal O}(150$ GeV$)$) and not too large (such that $m_\eta\lesssim {\cal O}(1.2$ TeV$)$). As discussed before, the exclusion becomes much weaker for a quasi-degenerate spectrum, and we quote the approximate limits obtained in \cite{Dreiner:2012gx} by reinterpreting CMS $\alpha_T$ and razor as well as ATLAS monojet analyses. The resulting limits are of the order of $m_\chi\gtrsim {\cal O}(200$ GeV$)$. A dedicated analysis of collider limits would definitely be desirable to obtain a more reliable estimate of the excluded regions.

\par The limits shown in Fig.~\ref{fig:mDMvsSplittingUR} demonstrate the complementarity of the various search strategies with respect to the mass range and the dependence on the mass splitting. In particular, direct detection experiments are very sensitive to (moderately) small mass splittings, while LHC limits appear to be most constraining for relative mass splittings of order one or larger. On the other hand, the range of dark matter masses probed by H.E.S.S.~lies above the present reach of direct detection and collider bounds. Nevertheless, H.E.S.S.~could exclude a thermally produced dark matter particle only if the annihilation signal from the central galactic halo was boosted by a factor ${\cal O}(10^2)$ relative to the flux resulting from an Einasto profile.

Naturally, one should keep in mind that these results apply only to the minimal mass-degenerate model discussed here. Some slight modifications of the model, like for example coupling to left-handed quarks or the presence of another scalar, should only have a minor impact and lead to qualitatively similar conclusions. However, a generalisation to less minimal models, which would clearly be interesting, reduces the predictivity and is beyond the scope of this work.   

\subsubsection{Prospects}

\par There is today a huge effort within the astroparticle physics community to improve our mapping of the gamma-ray sky with experiments such as H.E.S.S.~II \cite{hesssite}, MAGIC-II \cite{Tridon2010437}, GAMMA-400 \cite{Galper:2012fp}, DAMPE \cite{dampe} or CTA \cite{Consortium:2010bc}, and to push down the sensitivity of direct searches with the likes of CRESST \cite{cresstsite}, LUX \cite{luxsite}, XENON1T  \cite{Aprile:2012zx}, SuperCDMS \cite{supercdms}, EURECA \cite{eureca} or DARWIN \cite{Baudis:2012bc,darwin}. It is therefore timely to analyse the future prospects for constraining mass-degenerate scenarios through these two dark matter detection strategies.\FloatBarrier  We show in Fig.~\ref{fig:IB} (bottom) the expected 95\% C.L.~exclusion limits of GAMMA-400, CTA  and XENON1T for our three fiducial mass ratios $m_\eta/m_\chi=1.01,1.1,2$ and for couplings to up-quarks. Clearly, if no signal is observed, GAMMA-400 will provide the strongest constraints on the search for internal bremsstrahlung signatures up to $m_\chi\sim$200 GeV, whereas CTA will take over at higher masses with a flat dependence on the dark matter mass. The reason for this nearly constant sensitivity of CTA up to a few TeV is that the effective area grows rapidly with increasing energy, counteracting the usual dependence $m_\chi^{-2}$ due to the squared dark matter density. Another interesting feature of Fig.~\ref{fig:IB} (bottom) lies in the behaviour of the gamma-ray constraints with $m_\eta/m_\chi$. Similarly to the current limits discussed in the previous subsection, the projected bounds are only weakly sensitive to the actual value of the mass ratio and, for CTA, the sharper the IB spectral feature (i.e.~the smaller $m_\eta/m_\chi$), the stronger the constraint. However, the situation gets inverted in the case of GAMMA-400, for which the mass ratio $m_\eta/m_\chi=2$ leads to the most stringent limit. This is due to the extremely good energy resolution of GAMMA-400. In fact, with percent-level energy resolutions, one can start distinguishing the IB signature from the line at the end of the spectrum, the latter becoming more prominent at large mass splittings (as for $m_\eta/m_\chi=2$) and consequently driving the overall constraint. Note that the resolution featured by Fermi-LAT, H.E.S.S.~and CTA is not sufficient to separate internal bremsstrahlung from the line, and thus both contributions would show up as an extended spectral distortion in the measured gamma-ray spectrum. Also shown in Fig.~\ref{fig:IB} (bottom) are the XENON1T projected limits, which present the same behaviour with $m_\eta/m_\chi$ as the XENON100 limits drawn in Fig.~\ref{fig:IB} (top) and discussed in the previous subsection. XENON1T will boast impressive bounds on the models considered here (if dark matter couples to up-quarks), not only for small mass splittings but even for fairly non-degenerate configurations. For instance, in the case $m_\eta/m_\chi=2$, XENON1T shall overshadow gamma-ray constraints up to $m_\chi\sim 2-3$ TeV. Moreover, as discussed later on and illustrated in Fig.~\ref{fig:mDMvsSplittingUR} (bottom), the next generation of direct detection instruments will be able to probe thermal relics for a wide range of mass splittings and frequently up to TeV masses.

\par Fig.~\ref{fig:prospects} shows a summary of current and future constraints for the mass ratio $m_\eta/m_\chi=1.1$ and couplings to up- and bottom-quarks. In both cases, GAMMA-400 will improve upon Fermi-LAT (mainly due to its superior energy resolution) but not more than a factor of a few, whereas CTA will eventually supersede H.E.S.S.~by no more than one order of magnitude. We should caution at this point that our computation of gamma-ray prospects is conservative and can certainly be improved with a better knowledge of the actual instruments. Nevertheless, it should be appreciated that the present mapping of the gamma-ray sky provided by extremely precise experiments as Fermi-LAT, H.E.S.S.~and others leaves little room for orders-of-magnitude improvements.

\par Now, two striking results apparent from Fig.~\ref{fig:prospects} have important implications for the complementarity between direct and gamma-ray searches in the framework of mass-degenerate dark matter models. First, it is rather impressive that current limits from XENON100 exclude the possibility of observing spectral features with GAMMA-400 for couplings to the up- and bottom-quarks and a 10\% mass splitting. Such situation holds -- at least qualitatively -- for a large range of mass splittings and not only for the fiducial 10\% case shown in Fig.~\ref{fig:prospects}. Turning the argument around, if GAMMA-400 observes an IB spectral feature at any energy, that will constitute a strong indication that dark matter does not couple to (light) quarks.

\par The second striking result of Fig.~\ref{fig:prospects} is that the complementarity between CTA and XENON1T is close to maximal: the former will push down the sensitivity to spectral features in the whole mass range $m_\chi=\mathcal{O}(10)\textrm{ GeV}-\mathcal{O}(10)\textrm{ TeV}$ and with little dependence on the actual value of $m_\chi$, while the latter will probe extremely small cross sections but just up to a few TeV. Therefore, gamma-ray and direct searches really push the allowed parameter space in distinct, nearly orthogonal directions, implying good prospects to close in on mass-degenerate scenarios along the next decade. A glance at Fig.~\ref{fig:prospects} suggests that, for a non-negligible region of the parameter space, it even appears plausible that CTA and XENON1T find evidence for dark matter in a consistent way. This would be, of course, a truly exceptional situation with profound implications for dark matter model building, including a preference for mass-degenerate models with hadronic couplings.

\par Up to now we have analysed all the prospects in a global manner, with no particular focus on thermal production of WIMPs. Since the assumption of thermal relics shifts all the constraints in a non-trivial way, we present in Fig.~\ref{fig:mDMvsSplittingUR} (bottom) the thermal regions that will be probed by CTA and XENON1T. On the one hand, CTA shall not touch upon thermal cross sections (as also apparent from Figs.~\ref{fig:IB} and \ref{fig:prospects}) if the Einasto profile is an accurate description of the dark matter distribution towards the galactic centre. In case a boost $\gtrsim$10 is in order, CTA will cover an extensive mass range up to tens of TeV, particularly around 10\% mass splittings. On the other hand, XENON1T has the potential to test thermal candidates up to $\sim$1--2 TeV for $m_\eta/m_\chi=1.1$. This anticipates a neat complementarity between direct detection and gamma-ray searches in the near future. In a less optimistic note, assuming CTA and XENON1T spot no dark matter signal, the allowed parameter space of mass-degenerate models would be confined to the upper right corner in Fig.~\ref{fig:mDMvsSplittingUR} (bottom), namely $m_\chi\gtrsim1$ TeV, $m_\eta/m_\chi-1\gtrsim 30$\%. 

\par A certain portion of this corner may be probed by LHC after the upgrade of the centre-of-mass energy to $14$ TeV in searches for hadronic channels and missing transverse energy.  While detailed predictions for collider bounds require a dedicated analysis which is clearly beyond the scope of this work, it is safe to assume that in the most generic scenarios with sizable $m_{\eta}/m_{\chi}$ the limit on $m_{\eta}$ will be pushed into the multi-TeV region \cite{lhc14}. For a nearly mass-degenerate spectrum, the improvement in sensitivity depends on details of the analysis (such as jet energy cuts) and is affected by uncertainties related to e.g.~jet matching \cite{Dreiner:2012gx}. A considerable improvement for mass splittings in the range of a few tens of percent would potentially permit to probe thermal dark matter models with a light hadronic mediator at the LHC, and could also provide complementary information to XENON1T especially if dark matter couples to bottom quarks.

\begin{figure*}[t]
%\centering
\includegraphics[width=0.9\textwidth]{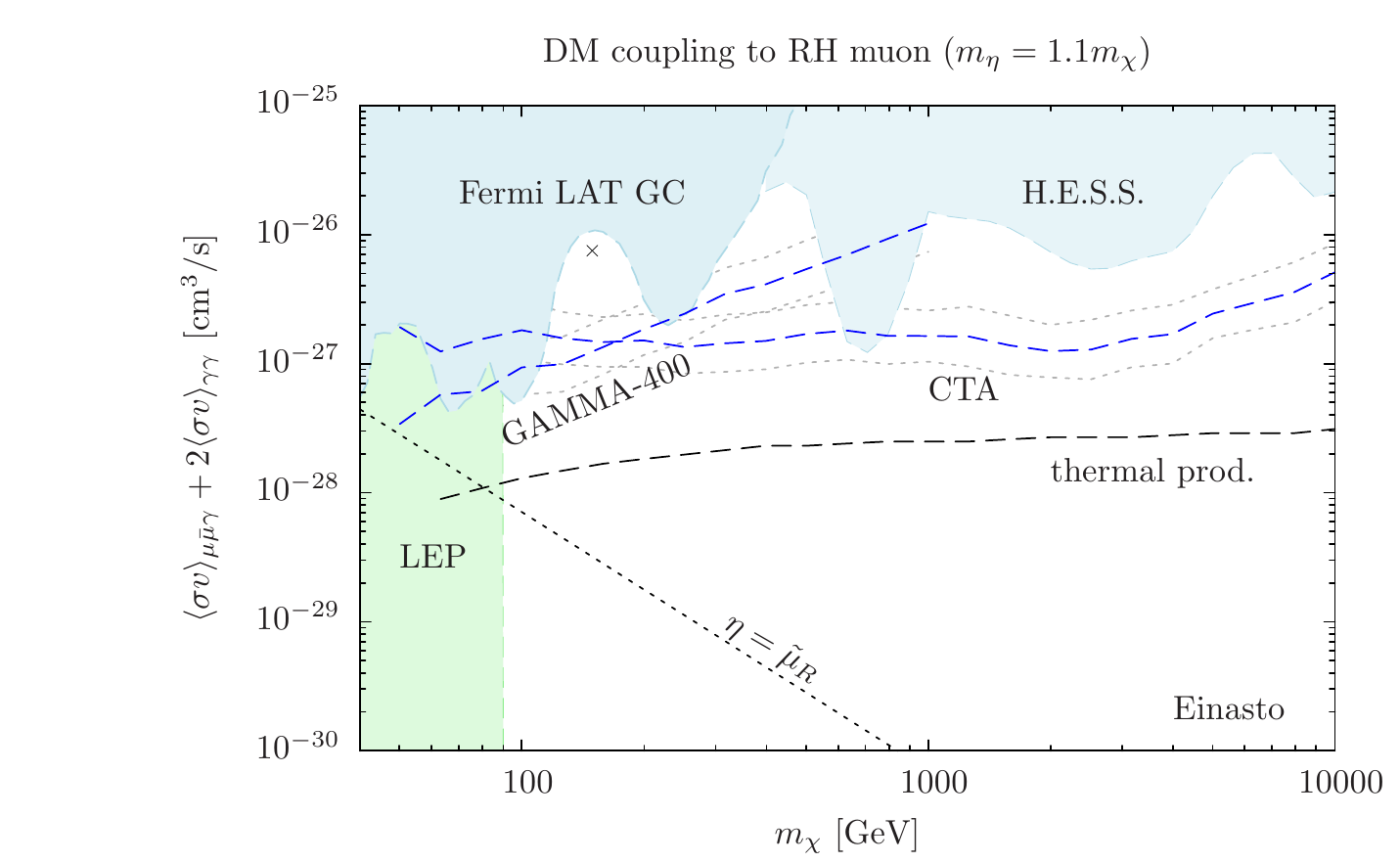}
\caption{As Figs.~\ref{fig:IB} and \ref{fig:prospects}, but for dark matter coupling to right-handed muons.} 
\label{fig:prospectsMuR}
\end{figure*}

\subsection{Coupling to leptons}

\begin{figure*}[htp]
\centering
\includegraphics[width=0.69\textwidth]{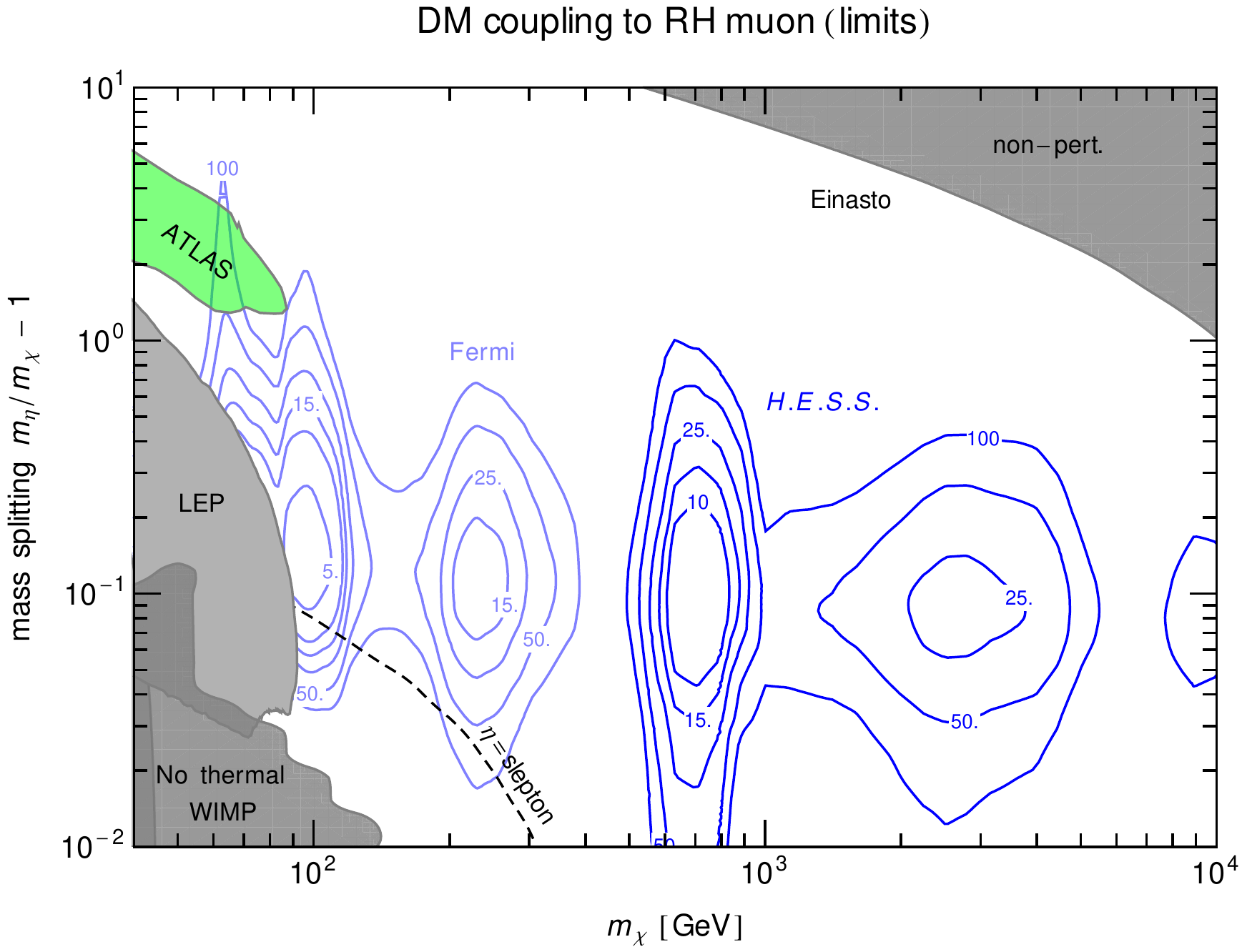}
\\[4ex]
\includegraphics[width=0.69\textwidth]{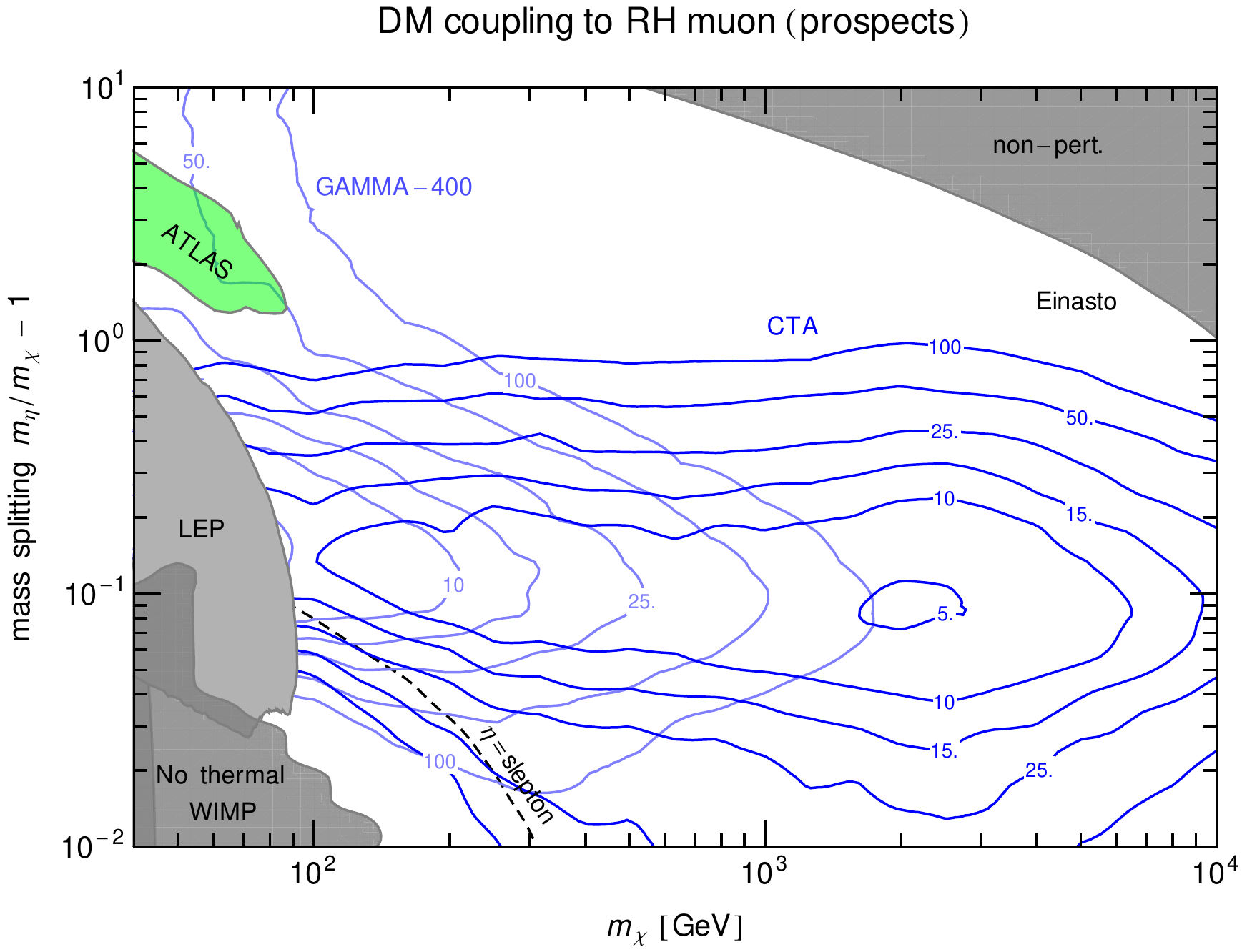}
\caption{As Fig.~\ref{fig:mDMvsSplittingUR}, but for dark matter coupling to right-handed muons.} 
\label{fig:mDMvsSplittingMuR}
\end{figure*}

\par The analysis of gamma-ray features in association to leptons, $\chi \chi \rightarrow l \bar l \gamma$, is completely analogous to the case of quarks, with the obvious substitution of the quark charge by the lepton charge. Taking for concreteness the case of a dark matter particle coupling to muons, the different charge of the muon compared to the quark leads to a small shift in the relative importance of the loop process $\gamma \gamma$ with respect to the $ \mu \bar \mu \gamma$ channel. However, for a small mass splitting $m_{\eta}/m_{\chi} =1.1$ the limits from Fermi-LAT and H.E.S.S.~remain essentially unchanged, as shown in Fig.~\ref{fig:prospectsMuR}. A more important difference arises from the limits from the observations of dwarf galaxies with the Fermi-LAT telescope. These limits are obtained from the non-observation of secondary gamma rays from dark matter annihilations, and therefore their strength depends on the ratio of secondary-to-primary gamma rays, which in turn depends substantially on the final state. In particular, the relative importance of the dwarf limits in the case of quarks mainly stems from the fact that the annihilation channels producing primary photons $q \bar q \gamma $ and $\gamma \gamma $ are always accompanied by the channels $q \bar q  g$ and $gg$  which, as seen in Fig.~\ref{fig:coupling}, have a significantly higher branching ratio and produce only secondary gamma rays. Consequently, these bounds are significantly weaker for leptons and are less stringent than the constraints from the searches for gamma-ray features \cite{Bringmann:2012vr}. Furthermore, in the case of couplings to leptons, direct detection experiments practically pose no limit, since the only interactions with the nucleus arise at the two loop level~\cite{Kopp:2009}. Besides, new physics that couples exclusively to leptons is inherently hard to probe at the LHC, making LEP limits still competitive at present. LEP II constraints require $m_{\eta} \gtrsim 90$ GeV  \cite{lepslepton} unless $m_{\eta}/m_{\chi} \lesssim 1.03$ as the searches lose sensitivity in the limit of degenerate masses. In that case the bound $m_{\eta} \gtrsim M_Z/2$ from the $Z$ decay width applies.

\par We show  in Fig.~\ref{fig:mDMvsSplittingMuR} the current status and prospects for this class of scenarios in the plane $m_{\chi}$ vs.  $m_{\eta} / m_{\chi}$, assuming that the dark matter particle was thermally produced. It follows from the figure that the sensitivity of gamma-ray observations reaches a maximum for $m_{\eta} / m_{\chi} \approx 1.1$,  for the same reason as discussed before for quarks. It is interesting to note that even with  cross sections of electroweak strength, there is a region in the parameter space where coannihilations are so efficient that it is not possible to produce thermally the whole cold dark matter population. 

\par Searches for charged colourless scalars at hadron colliders rely mainly on Drell-Yan production and are therefore practically independent of the size of the coupling constant $f$. LHC searches for muon pairs and missing energy  \cite{atlasslepton} are starting to probe the parameter space at low masses and high splittings, whereas the LEP limits \cite{lepslepton} still stand as the strongest ones in the region with low splittings. The ATLAS limits depend rather strongly on the leptonic flavour since taus decay mostly hadronically, thus making their identification rather challenging. In general, collider searches do not probe $m_{\chi} \gtrsim 100$ GeV so far, therefore gamma-ray observations remain the most sensitive probe for dark matter coupling to leptons. The next generation of gamma-ray telescopes like GAMMA-400 and CTA will continue closing in on this class of scenarios, see Fig.~\ref{fig:mDMvsSplittingMuR} (bottom); however, thermally produced dark matter particles might easily escape detection, unless the rate of annihilations is enhanced by astrophysical or particle physics boost factors.

\FloatBarrier
\section{Conclusions}\label{conclusions}

\par The complementarity between direct, indirect and collider searches for dark matter is a relatively new possibility and likely to acquire increasing importance in the coming years. Given the absence of evidence for full-fledged theoretical frameworks (such as supersymmetry) from the LHC or elsewhere, minimal models -- that very often predict spectacular signatures -- have become rather popular lately. In this work, we constructed minimal mass-degenerate toy models and studied comprehensively the corresponding internal bremsstrahlung constraints given the current status of direct searches. It turns out that, despite impressively stringent and displaying an almost flat trend across the GeV--TeV mass range, the latest internal bremsstrahlung limits are still far from probing thermal relics. In the future, if no spectral feature is detected, GAMMA-400 and CTA will improve significantly upon the existing bounds at low and high masses, respectively. However, we conclude that both experiments will fall short of the thermal region in the models considered for both hadronic and leptonic couplings, unless the boost factor towards the galactic centre is in excess of a factor of a few. Instead, XENON100 excludes already thermal candidates at intermediate mass splittings up to hundreds  of GeV, if dark matter couples to one of the light quarks. Incidentally, current and upcoming direct searches have crucial implications on the planning of future internal bremsstrahlung campaigns. Firstly, we find that for minimal models with pure couplings to up-quarks a gamma-ray space telescope sensitive up to the TeV with superior energy resolution -- such as GAMMA-400 -- shall only cover a portion of the parameter space already excluded by the latest XENON100 data. Secondly, the prospects for CTA and XENON1T are particularly complementary and it is even possible that each experiment will provide independent, compatible evidence for dark matter. This would constitute a suggestive argument in favour of mass-degenerate scenarios. Otherwise, if no signal is observed by CTA nor XENON1T, then minimal mass-degenerate models with couplings to light quarks will be confined to large dark matter masses and large mass splittings. Closing that gap will be extremely challenging even on the long run.

\vspace{0.5cm}
{\it Acknowledgements:} The authors are grateful to Torsten Bringmann and Laura Lopez-Honorez for useful comments. This work has been partially supported by the DFG cluster of excellence ``Origin and Structure of the Universe'' and by the DFG Collaborative Research Center 676 ``Particles, Strings and the Early Universe''. A.I.~and M.P.~would like to thank the Kavli Institute for Theoretical Physics in Santa Barbara, California, for hospitality during the final stages of this work. S.V.~also acknowledges support from the DFG Graduiertenkolleg ``Particle Physics at the Energy Frontier of New Phenomena''.

%%%%%%%%%%%%%%%%%

\bibliographystyle{JHEP}

\bibliography{DDvsGamma}

\end{document}